\documentclass[iop,onecolumn]{revtex4}

\usepackage{amsmath}
\usepackage{amssymb}
\usepackage{braket}
\usepackage{hyperref}
\usepackage[usenames,dvipsnames]{color}

\usepackage{graphicx}
\usepackage[tight,footnotesize]{subfigure}

\usepackage{tikz}
\usetikzlibrary{tikzmark}
\usetikzlibrary{shapes,shadows,arrows}
\usetikzlibrary{decorations.markings,decorations.pathmorphing}
\usepackage{hyperref}

\DeclareMathOperator{\Tr}{Tr}

\newcommand{\ct}{\cite}

\newcommand{\bi}{\bibitem}
\newcommand{\be}{\begin{equation}}
\newcommand{\ee}{\end{equation}}
\newcommand{\ba}{\begin{eqnarray}}
\newcommand{\ea}{\end{eqnarray}}
\newcommand{\al}{\alpha}
\newcommand{\ep}{\epsilon}

\newcommand{\non}{\nonumber}

\newcommand{\noi}{\noindent}
\newcommand{\om}{\omega}

\begin{document}

\title{One-dimensional quantum many body systems with long-range interactions}
\author{Somnath Maity}
\email{msomnath@iitk.ac.in}
\author{Utso Bhattacharya}
\author{Amit Dutta}
\affiliation{Department of Physics, Indian Institute of Technology Kanpur, Kanpur 208016, India}

\begin{abstract}

The presence of algebraically decaying long-range interactions may alter the critical as well as topological behaviour of a quantum many-body systems. However, when the interaction decays
at a faster rate, the short-range behaviour is expected to be retrieved. Similarly, the long-range nature of interactions has a prominent signature on the out of equilibrium dynamics of these systems,
e.g,  in the growth of the entanglement entropy following a quench, the propagation of mutual information and non-equilibrium phase transitions. In this review, we summarize the results
of long-range interacting classical and quantum Ising chains mentioning some recent results. Thereafter, we focus on the recent developments on the integrable long-range Kitaev
chain emphasising the role of long-range superconducting pairing term on determining its topological phase diagram and out of equilibrium dynamics and also incorporate relevant discussions
on the corresponding Ising version.

\end{abstract}
\maketitle

\section{Introduction}

The critical behavior of a classical as well as a quantum system is drastically modified in the presence of a long-range interaction that decays in a power-law fashion. The study
of long-range interacting Ising (as well as $n$-vector systems) has a long history\ct{rulle68,dyson69,kac69}. The model was studied in the context of possibility of
long-range order surviving at a finite temperature in the  one-dimensional situation\ct{thouless69}.  However for all dimensions, the critical behavior of  the same model  was explored 
 within the renormalisation group 
technique. Remarkably, it was also found that one of the critical exponents is exactly known as long as the long-range interactions are relevant \ct{fisher72}.  Further, if the strength of the interaction decays 
 fast enough (with the distance between the spins),  the model exhibits the short-range critical behaviour. Many magnetic systems
show the signature of long-range interactions (e.g., RKKY interactions); the celebrated example being the  experimental studies of the 
 Ising system ${\rm LiHo{_x}Y_{1-x}F_4}$ which has  long-ranged dipolar interactions and can be ideally modelled by a quantum Ising spin-glass with long-range
 interactions \ct{wu91}. Recently, such tunable power-law interaction has been realized in trapped
ions \ct{britton12, islam13,richemere14}, thereby providing an excellent platform for verifying  the corresponding theoretical predictions. In this review, we present a brief summary of long-range interacting classical and quantum Ising models elaborating on how do the the critical behaviour of the model gets modified
in the presence of long-range interactions; a special type of Brezenski-Kosterlitz-Thouless transition occurring in an inverse-square classical Ising chain has also
been discussed. We also briefly dwell on the quantum version of the long-range Ising chain that has a non-commuting transverse field and discuss the associated zero
temperature and finite temperature phase transitions.

Topological properties of a $p$-wave spin-less superconductor (i.e., the short-range Kitaev chain \ct{kitaev01}), have been studied from
different points of views in recent years \ct{fulga11,sau12,lutchyn11,degottardi11,degottardi13,thakurathi13,degottardi13_1}. For a review, see [\onlinecite{alicea12}].
 A spinless p-wave superconductor may have  a phase with edge Majorana modes as midgap
excitations that are guaranteed by particle-hole symmetry of the system. The superconducting pairing term
of such a Hamiltonian actually induces the zero energy
excitations in the system. The presence  of the
zero energy Majorana modes at the ends of a long and
open chain characterises topologically non-trivial phase;  on the contrary, the phase in which edge Majoranas are absent  is  topologically trivial. The number of Majorana zero modes (MZMs)
is a topological invariant; this value does not change until the system crosses the phase boundary between the
topological and the trivial phases. It has been proposed that
the proximity effect between the surface states of a strong
topological insulator and s-wave superconductor can generate a two dimensional state strongly resembling a p-wave superconductor which can host Majorana states
at the vortices \ct{fu08}. Further, topological superconductivity can be induced by replacing the topological insulator by a semi-conductor with strong
spin-orbit coupling \ct{sau12}. Recent experiments have reported possible detection of the
signature of these Majorana modes in the zero-bias transport
properties of nanowires coupled to superconductors
\ct{mourik12,deng12,das12,chang13}.
An experimental study of the possible hybridization of the  Majorana modes (and hence their disappearance) by tuning the chemical potential of
a similar system across a topological phase transition has been carried out \ct{flink13}.
However,  some discrepancies in experimental results
with the theoretical predictions have also been pointed out \ct{rainis13}. We note that the short-range Kitaev chain can be mapped to a quantum Ising chain through the Jordan-Wigner 
transformation \ct{sachdev10,lieb64}, however the latter model does not have any topology in the true sense.

Recently motivated by the short-range one-dimensional
Kitaev chain \ct{kitaev01}, a long-range version of an integrable p-wave superconducting chain of fermions, with a long-range
superconducting pairing term was proposed \ct{vodola14} (see also \onlinecite{vodola_th,huang14,viyuela16,vodola16,viyuela15,lepori17,regemortel16,lepori171,giuliano18,cats18}).  
A major part of this review deals with the  static as well as the dynamical properties of this long-range Kitaev (LRK) chain. What  is remarkable is that in spite of the power-law interacting
superconducting term,  this model can be reduced to decoupled  $2 \times 2$ structure corresponding to each
quasi-momentum value and hence is {\it exactly solvable}. It should, however, be noted that unlike the short-range Kitaev (SRK) model, this model can not be mapped to a corresponding spin model. It has been reported  \ct{viyuela16} that when
the pairing terms decay faster, the model exhibits the topological properties of the
corresponding short-range  model; on the contrary, for slow
decay of the long-range interactions  the model
supports a new unconventional topological phase of matter.
In this new phase, the zero-energy Majorana modes coalesce to form
massive non-local edge states called massive Dirac modes (MDMs)
which are otherwise absent in the standard Kitaev model.
These new edge states lie within the bulk energy gap and are
topologically protected against local perturbations that do not
break fermionic parity and particle-hole symmetry.

Unlike the short-range 
case, the LRK and Ising chains are two independent models.  The
LRK chain is integrable while  the long-range  Ising chain is not, and hence
two models can not be mapped to each other.   The emphasis in this review is more on   the LRK chain; this is because of its integrable nature which has led to rigorous  analytical studies.  On the other hand, for the Ising chain, the results (including the phase diagram) are obtained mostly using numerical methods.
However, in every section  where we have discussed the LRK chains, the corresponding  results
of  the long-range Ising chain have been mentioned with appropriate references.

The review is organised in the following fashion: In Sec. \ref{sec_LRQI}, we discuss phase transitions in  long-range interacting classical and quantum Ising chains. In Sec. \ref{sec_LRK}, we
introduce the LRK Hamiltonian and provide an exact analytical solution of the spectrum. In Sec.\ref{srk}, we recall the short-range limit and analyse the corresponding topological phase
diagram. The critical points, entanglement entropy and the topological phase diagram of the  LRK model are presented in Sec. \ref{sec_topology}. In Sec. \ref{sec_lrk_dynamics}, we review the
out of equilibrium dynamics of the LRK chain in the context of propagation of mutual information and the Kibble Zurek scaling and some relevant studies on the dynamics of the LRK chain
are briefly mentioned in Sec. \ref{sec_related}. Finally, we present the concluding comments in Sec. \ref{sec_conclusion}. In Appendix A, we present a brief calculation of the winding number
of the LRK chain both in short-range and long-range limits.

\section{Long-range interacting classical and  quantum Ising models}
\label{sec_LRQI}

Let us first revisit the long-range  interacting classical Ising model with (ferromagnetic) interactions decaying in an algebraic fashion. The model is described by the Hamiltonian

\be
H = - \sum_{(i,j)} J_{ij} S_i S_j,
\label{eq_ham}
\ee
where the binary variables $S_i$'s, residing on the site $i$, can assume values $\pm 1$. The interaction is chosen of the form: $$J_{ij} = J/r_{ij}^{d+\sigma},$$
where $r_{ij}$ is the distance between the sites $i$ and $j$ and $d$ denotes the spatial dimension of the hyper-cubic lattice. Here, the parameter $\sigma$ determines the strength of interaction
between the distant sites and  for a large value of $ \sigma$, the model reduces to the short-range nearest neighbour Ising model. However, we shall discuss below that if the range parameter $\sigma$ exceeds $2$ 
(for $d>1$), the critical behavior of the model is essentially that of the short-range system. The model in Eq.~\eqref{eq_ham} is not exactly solvable, except for the special limit of the nearest neighbour
interaction in one and two dimensions.

The study of a power-law interacting classical Ising model has a long history \ct{rulle68,dyson69,kac69}. In
particluar, Thouless \ct{thouless69} explored the possibility of a long-range ferromagnetic order surviving at a finite temperature in one dimensional version of the model in Eq.~\eqref{eq_ham}. It was
argued that with $d=1$, a finite critical temperature, demarcating the transition between the ferromagnetic and the paramagnetic phase, can only exist if the parameter $\sigma <1$, i.e., the interaction is sufficiently long-ranged. For $\sigma >1$, however, the long-range 
order gets destroyed at any finite temperature. (This can also be verified in a straightforward fashion using a generalized version of the Peierls' argument extended to the long-range interacting
systems.)

\begin{figure}[]
	\centering 
	\tikzstyle{arrow1} = [thick, ->,>=stealth]
	\tikzstyle{arrow2} = [ <->,>=stealth]
	\tikzstyle{line} = [dashed,thick,-,>=stealth]
	\tikzstyle{line1} = [-,red!80!black]
	\begin{tikzpicture}   
	\draw[arrow1] (-3.5,0) -- node[below=8pt,right=50pt,fill=white] {$T_c$} (3.5,0) node[below=8pt,left=3pt,fill=white] {$T$};  
	\draw[line] (0,0)node[above=40pt,fill=white] {\small{Ordered Phase}} -- node[below=3pt,fill=white] {\small{correlation $\sim \frac{1}{r^\theta}$}} (0,2.5)node[above]{$\theta=1$}node[above=25pt,fill=white] {$1D$ classical chain ($\sigma=1$)};
	\draw[line] (2,0) node[above=35pt,right=3pt,fill=white] {\small{Disordered Phase}} -- (2,2.5)node[above]{$\theta=0$};
	\draw[line] (-2,0) node[above=35pt,left=3pt,fill=white] {\small{Ordered Phase}} --node[below=12pt,left=5pt,fill=white] {\small{correlation $\sim \frac{1}{r^2}$}} (-2,2.5)node[above]{$\theta=2$};
	\draw[line1] (-0.1,-0.1) -- (0.1,0.1);
	\draw[line1] (0,-0.1) -- (0.2,0.1);
	\draw[line1] (0.1,-0.1) -- (0.3,0.1);
	\draw[line1] (0.2,-0.1) -- (0.4,0.1);
	\draw[line1] (0.3,-0.1) -- (0.5,0.1);
	\draw[line1] (0.4,-0.1) -- (0.6,0.1);
	\draw[line1] (0.5,-0.1) -- (0.7,0.1);
	\draw[line1] (0.6,-0.1) -- (0.8,0.1);
	\draw[line1] (0.7,-0.1) -- (0.9,0.1);
	\draw[line1] (0.8,-0.1) -- (1.0,0.1);
	\draw[line1] (0.9,-0.1) -- (1.1,0.1);
	\draw[line1] (1.0,-0.1) -- (1.2,0.1);
	\draw[line1] (1.1,-0.1) -- (1.3,0.1);
	\draw[line1] (1.2,-0.1) -- (1.4,0.1);
	\draw[line1] (1.3,-0.1) -- (1.5,0.1);
	\draw[line1] (1.4,-0.1) -- (1.6,0.1);
	\draw[line1] (1.5,-0.1) -- (1.7,0.1);
	\draw[line1] (1.6,-0.1) -- (1.8,0.1);
	\draw[line1] (1.7,-0.1) -- (1.9,0.1);
	\draw[line1] (1.8,-0.1) -- (2.0,0.1);
	\draw[line1] (1.9,-0.1) -- (2.1,0.1);
	\draw[fill=black] (2,0) circle (.3ex);
	\end{tikzpicture}
	
	\caption{The schematic phase diagram of the inverse-square Ising chain; Hamiltonian \eqref{eq_ham} with $d=\sigma=1$; the exponent varies continuously in the ordered phase and
		the linear susceptibility diverges in the shaded region. After Ref. [\onlinecite{dutta03}].}
	\label{classical_dia}
\end{figure}
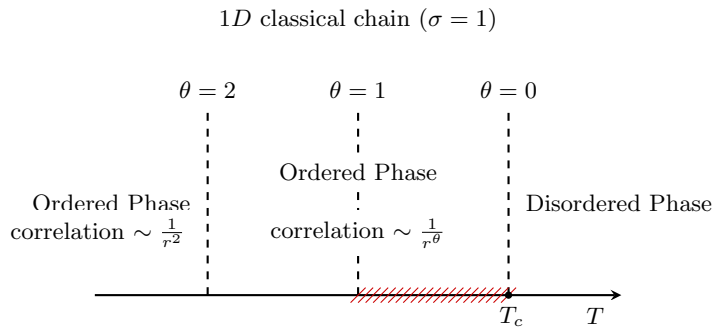

The situation with $d=\sigma=1$ (i.e., the inverse-square classical Ising model), which happens to be the marginal case   turns out to be the most intriguing.  This model exhibits a special kind
of Brezenski-Kosterlitz-Thouless (BKT) transition at a finite temperature $T_c$ \ct{kosterlitz76}. This transition is  associated with a  discontinuous jump in the spontaneous magnetisation; this resembles the discontinuous
jump in the spin-wave stiffness in the conventional BKT transition occurring in a two-dimensional classical XY model. The kinks and anti-kinks in the present case interact logarithmically
and a systematic decimation of the nearest neighbour kink-antikink pairs yields the BKT RG equation. Referring to a Coulomb gas picture, the present model describes a two-dimensional
Coulomb gas (due to  logarithmic interactions) confined in one-dimension with the constraint that kinks and anti-kinks are alternating. The interesting feature of the model is the existence
of a phase (as predicted  for the first time in Ref. [\onlinecite{bhattacharjee81}]) characterised by a   correlation function $G(r)$ (between two spins separated by a distance $r$) falling   as $ G(r) \sim 1/r^{\theta}$.  Near the transition point the exponent 
$\theta$ decreases with temperature and vanishes at $T_c$ where there is a logarithmic decay of correlation function. Within the range of temperature, for which $\theta$ lies between 0 and 1, the linear susceptibility of the system
diverges as schematically depicted in Fig.~\ref{classical_dia}. The existence of
this power-law correlated phase and the  discontinuous jump in the order parameter  were established using the renormalisation group arguments along with  numerical simulations
(up to $10^6$ lattice sites) a couple of decades after the initial prediction \ct{luijten01}.
This situation may be contrasted with the conventional XY model where in the entire BKT phase, the exponent describing the algebraic
decay of the correlation function varies continuously and saturates to a value $1/4$ at the transition; further, the linear susceptibility diverges for all  temperature $T < T_c$ \ct{chaikin,goldenfeld92}.  It should
also be noted that an inverse square Ising chain has been studied in the context of  the Kondo effect in a metal \ct{anderson71}.

The field theoretical study of the associated BKT transition in the inverse-square model is still an {\it open} question. Even though a one-dimensional sine-Gordon action (with a long-range
kinetic term) indicates a transition, the coupling term (or the inverse temperature) does not flow to any order of perturbation theory \ct{narayan99}. This sine-Gordon action does not correctly  represent
the BKT transition of the inverse square chain because it fails to incorporate the constraint of alternating kink-antikink pairs.

The model in Eq.~\eqref{eq_ham} has been studied using a renormalisation calculation  using an $\epsilon$ expansion around the upper critical dimension \ct{fisher72}.
%
%
It has been concluded that for $\sigma <2$, the long-ranged nature of the interaction dominates while for $\sigma \geq 2$, the critical
exponents are those of the short-range Ising model. 
It is also straightforward to show that 
for a given $\sigma$ the upper-critical dimension of the model $d_u =2\sigma$, which approaches the well-known result  $d_u=4$ in the short-range limit of
$\sigma \to 2$. The mean-field exponents for the long-range model are simply given as $\gamma_{\rm LR}=1$, $\nu_{\rm LR}=1/\sigma$ and the Fisher exponent (that determines the algebraic
decay of the correlation function at the critical point) $\eta_{\rm LR}=2-\sigma$;  evidently all
the exponents approach the short-range mean field exponents as $\sigma \to 2$. 

We further note that the upper-critcality condition, $d_u = 2 \sigma$, also determines an upper-critical range beyond which mean field critical behaviour holds true. For example, for
$d=1$, for the parameter $\sigma <1/2$ (i.e., the interaction is sufficiently long-ranged), we expect the critical exponents are those of mean field character; on the contrary, for $1>\sigma >1/2$, there
will be {\it non-meanfield} exponents which can be derived using the $\epsilon$-expansion technique around the upper critical dimension. What is remarkable is that the exponent 
$\eta$ sticks to the mean field value $2 -\sigma$ up to any order  of $\epsilon$ as long as $\sigma <2$. As $\sigma \to 2$, $\eta_{\rm LR}$ picks up the value of the  short-range Ising model ($\eta_{\rm SR} = \epsilon^2/54$ where $\epsilon =4-d$).
(In that sense, one can argue that more precisely, the crossover between the long-range and
the short-range critical behaviour occurs when $ \sigma \to 2 -\eta_{\rm SR}$.) Therefore, one of the two independent exponents associated
with the continuous phase transition is exactly known.

The quantum version of the model in Eq.~ \eqref{eq_ham} described by the Hamiltonian
\be
H = -  \sum_{(i,j)} J_{ij} \sigma_i^z \sigma^z_j -h  \sum_i\sigma_i^x,
\label{eq_ham2}
\ee
was introduced in the Ref. \cite{dutta01}; here, $\sigma$'s are Pauli matrices satisfying the standard commutation relations. The non-commuting transverse field $h$ can be
tuned to drive the system from a ferromagnetic phase to a paramagnetic phase across a quantum critical point (QCP) at $T=0$. 
%
It was established that for $\sigma <2$, the quantum critical behavior is determined by the long-ranged nature of the interaction. The mean field dynamical exponent ($z$) that determines
the dispersion at the quantum critical point (i.e., $\om \sim q^z$)  is given by $z = \sigma/2$ and thus assumes the short-range value $z=1$ as $\sigma \to 2$. Using
the corresponding bare propagator, one can deduce the modified upper criticality relation $d_u =3\sigma/2$. One can further show that the dynamical exponent
$z$ picks up a correction only in the $O(\epsilon^2)$. Exploiting the modified upper criticality relation, we find that for a given $d$, the quantum transitions with values $\sigma \leq 2d/3$ are described
by the $\sigma$-dependent mean field exponents quoted above.

We shall now elaborate on the one-dimensional situation: what is noteworthy is that the Hamiltonian \eqref{eq_ham2} with $d=1$ shows a quantum phase transition for all values of $\sigma$
unlike the classical case when the long-range order gets destroyed at any finite temperature for $\sigma >1$. This is because of the quantum-classical mapping of the Hamiltonian
\eqref{eq_ham2}: the quantum 
phase transition of  the one-dimensional
version of the quantum model in \eqref{eq_ham2} at $T=0$,  belongs to the same universality class as that of the finite temperature phase transition of a two-dimensional classical Ising model. This equivalent
classical model has  a power-law interaction in one direction and short-range interaction in the other (Trotter) direction and shows a finite temperature phase
transition even in the short-range limit.  However, any  finite temperature transition of the same quantum model belongs to the universality class of an equivalent classical Ising model of
same dimensionality. Thus,  for  $\sigma >1$, there is no finite temperature transition of the model Eq.~\eqref{eq_ham2}, while there is a phase transition line separating the ferromagnetic
and the paramagnetic phase in the $T$-$h$ plane if $\sigma <1$. 

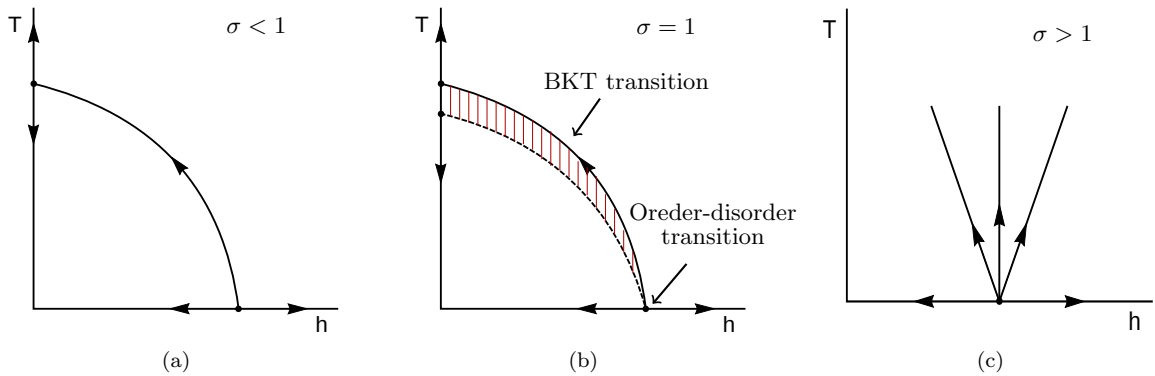
\begin{figure}
	\centering
	\subfigure[]{
		\includegraphics[width=0.25\columnwidth,height=0.25\columnwidth]{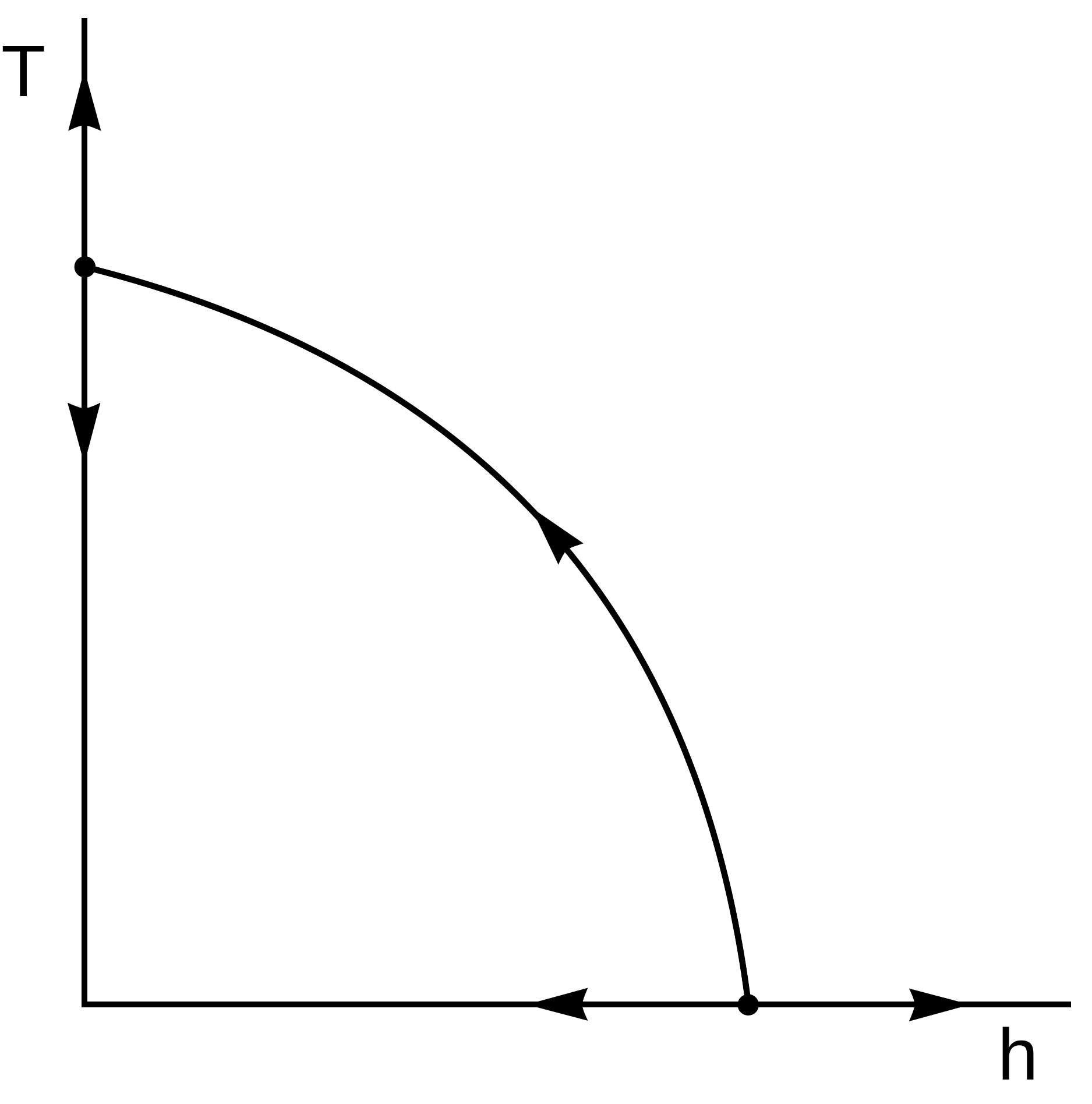}
		\label{quantum1}}
	\hspace{0.5cm}
	\subfigure[]{
		\includegraphics[width=0.25\columnwidth,height=0.25\columnwidth]{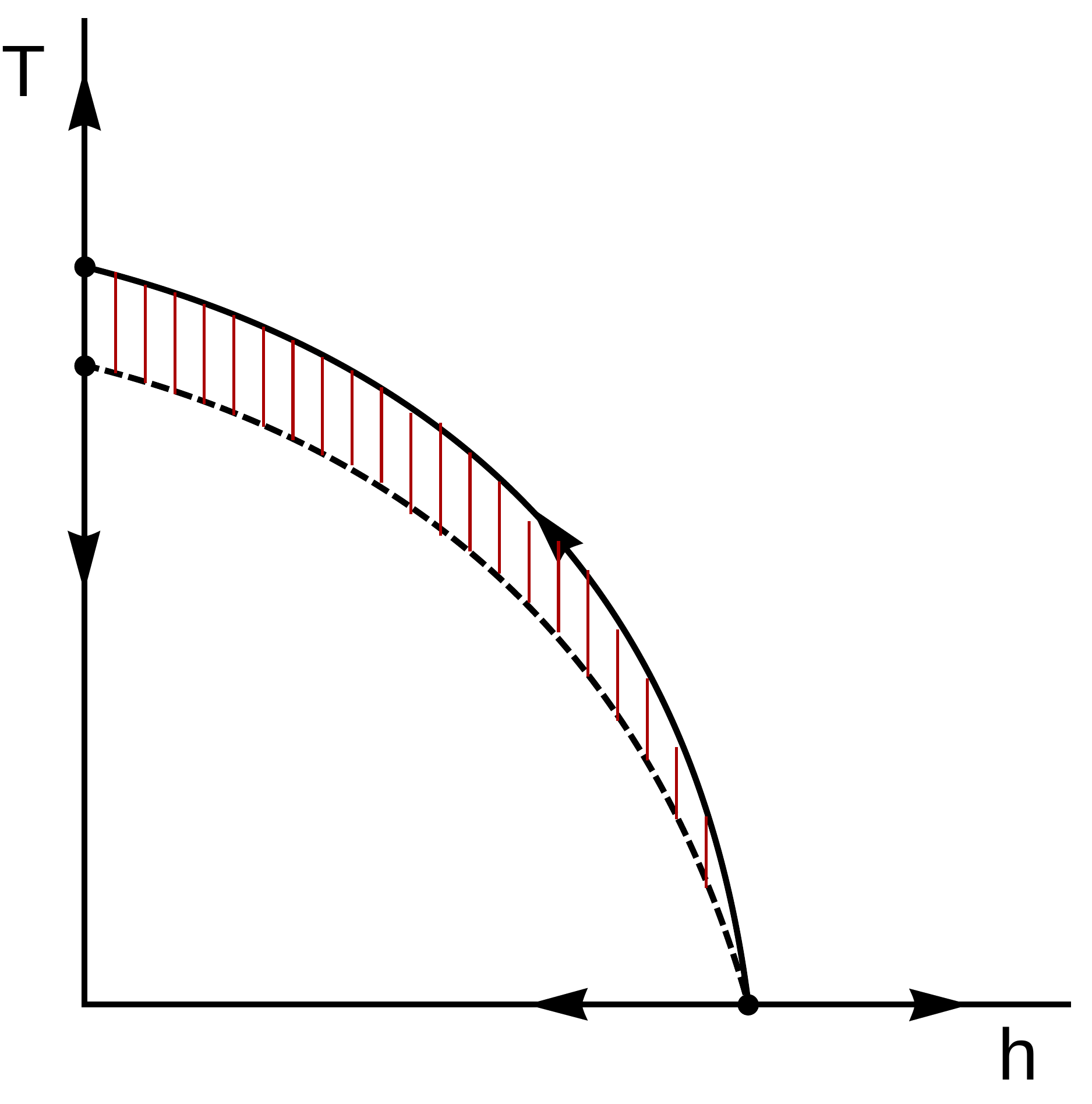}
		\label{quantum2}}
	\hspace{0.5cm}
	\subfigure[]{
		\includegraphics[width=0.25\columnwidth,height=0.25\columnwidth]{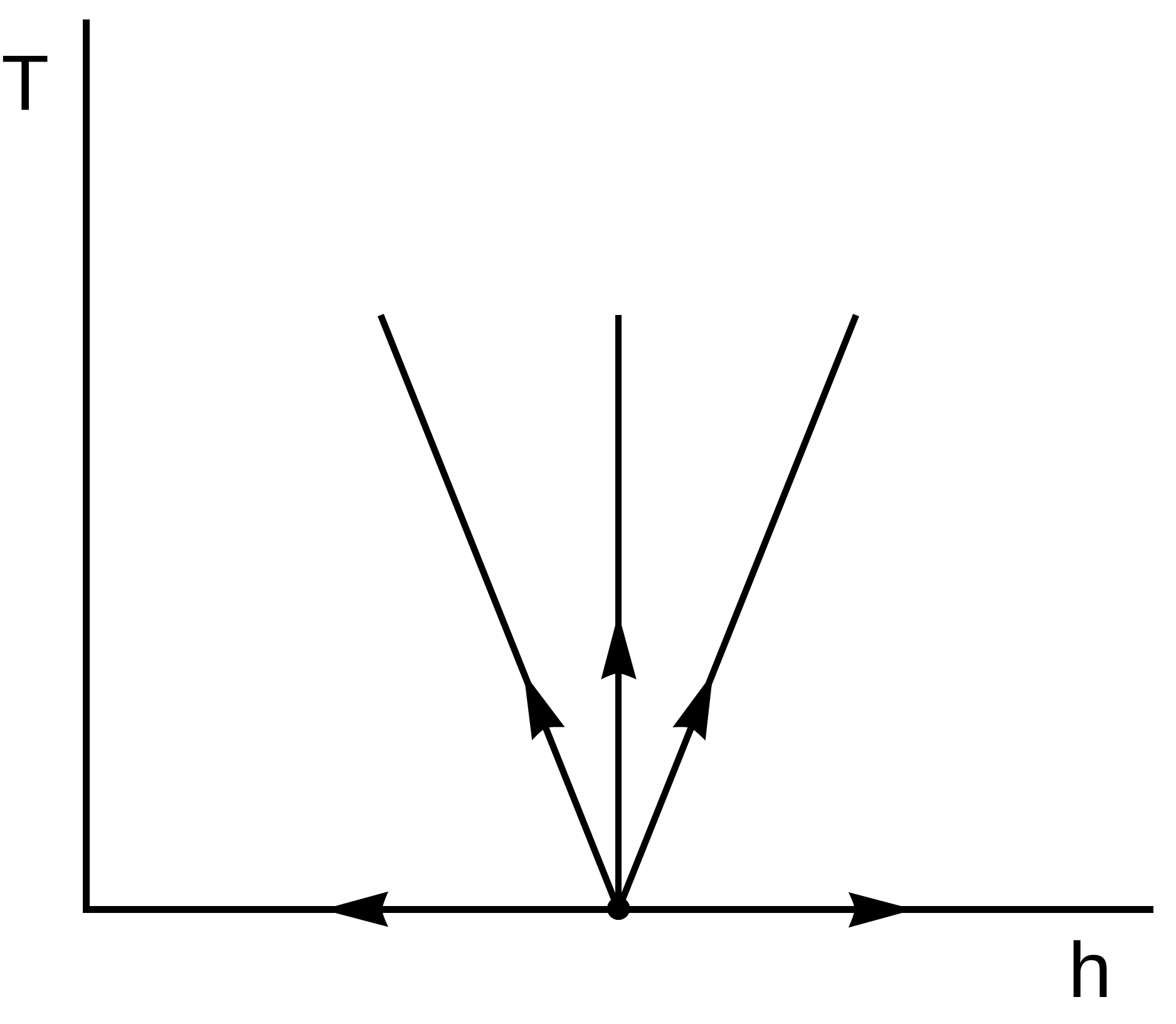}
		\label{quantum3}}
	\begin{tikzpicture}[remember picture, overlay]
	\draw [thick,<-] (-6.95,0.47) -- (-6.5,1.0) node [black,right=10pt,above=13pt] {\small {Oreder-disorder}}node [black,right=10pt,above=4pt] {\small {transition}};
	\draw [thick,<-]  (-8.0,2.7) node[above=60pt]{$1D$ quantum Ising chain} node[right=35pt,above=35pt]{$\sigma=1$} node[left=120pt,above=35pt]{$\sigma<1$} -- node[right=180pt,above=25pt]{$\sigma>1$} (-7.65,3.2) node [black,right=10pt,above=0pt] {\small {BKT transition}};
	\end{tikzpicture}
	\caption{The schematic phase diagram and the RG flow in the $h$-$T$ plane of one dimensional quantum Ising chain \eqref{eq_ham2} for different values of $\sigma$. (a)  For $\sigma <1$, the system
		has both thermal phase transitions and  a quantum phase transition (at $T=0$) and there is a phase boundary separating the ordered and the disordered phase.  The finite temperature transitions
		belong to the classical long-range universality class. (b) For $\sigma=1$, any finite temperature transition is of BKT type while the zero-temperature transition is order-disorder. In the shaded
		region, linear susceptibility diverges. (c) For $\sigma >1$, there is no finite temperature transition. (After Ref. [\onlinecite{dutta02}].) }
	\label{quantum_dia}
\end{figure}

We note that the disordered version of the  long-range Ising model has also been studied in the context of classical \ct{kotliar83} as well as
quantum  spin glass transitions \ct{dutta02}. The phase transition of a short-range Ising model on a long-range percolating lattice \ct{aizenmann88} has also
been explored \ct{dutta03}.

We shall now focus on the case $d=\sigma=1$, where, as discussed already, the classical version of the model shows a special type of BKT transition. Recalling the quantum-classical
mapping, we note that the finite temperature transition of the quantum Ising chain is always determined by the thermal fixed point. We can thus conclude that any finite temperature
transition (for $\sigma=1$) is definitely a BKT transition occurring at a reduced temperature $T_c(h)$. Furthermore, for $\sigma >1$, the model behaves like the short-range Ising
chain at any finite temperature and there is no long-range order (see Fig~\ref{quantum_dia}).

Let us now briefly dwell  on some recent studies: the fidelity susceptibility of quantum antiferromagnetic transverse Ising chain with a long-range
power law interaction $1/r^{\alpha}$, has been studied
using the large-scale density matrix renormalization group method \ct{sun17};  the complete phase diagram has been obtained for $0 < \alpha  \leq 3$ from the data
collapse of the fidelity susceptibility . It has been reported that the critical exponent of the correlation length  
$\nu = 1$ for arbitrary $\alpha  > 0$, indicating that  quantum phase transitions for all values of $\alpha$ are of second-order Ising nature; further,
the critical value of the transverse field  $h_c$ has been found to  change monotonically with
respect to $\alpha$; the same model has been studied using the linked cluster method \ct{fey16}.

We conclude this section with the comment that recently, a general  $d$ ($\geq2$) dimensional classical continuous symmetric $O(N)$ model with long-range interactions decaying as a power law with exponent $(d+\sigma)$ has been studied using functional renormalization group methods~\cite{defenu15}. Interestingly, in this case, the critical exponents of the long-range $O(N)$ models can be found from the corresponding short-range $O(N)$ models with an effective fractional dimension. Furthermore, by varying $\sigma$ while keeping $d$ fixed, a sequence of different multicritical long-range universality classes appear; this is  analogous to the upper critical dimension for the short-range models when $d$ is varied. The phase diagram and the critical exponents of anisotropic long-range spin systems has also been studied~\cite{defenu16}. Further,
the three dimensional long-range Ising model has been studied using the conformal bootstrap  analysis \ct{behan19}.   The phases and criticality of the $O(N)$ symmetric $d$ dimensional quantum rotor model \ct{dutta01} with long-range interaction has also been studied using the functional renormalization group techniques in order to find the connection between quantum long-range systems and the classical long-range systems~\cite{defenu17}. It has been reported that the mapping of long-range quantum systems to a corresponding long-range classical universality class is only possible for large value of $\sigma$ but this not the case for $\sigma$ less than a critical value when the correlation functions become highly anisotropic in both spatial and temporal coordinates.

\section{Long-range Kitaev model}
\label{sec_LRK}

In this section, we will consider another example of long-range quantum system related to spinless fermionic models which is a generalised version of the $p$-wave superconducting Kitaev chain~\cite{kitaev01}, where both the hopping and $p$-wave pairing terms may have long range interactions. Unlike the long-range Ising model, discussed in the previous section, the presence of long-range interactions  does not break the integrability of the model.  The long-range Kitaev chain,  described by the tight-binding Hamiltonian,
\ba
H_{LRK}=\sum_{n=1}^{L}-w \left(c_n^{\dagger}c_{n+1}+c_{n+1}^{\dagger}c_n\right)-\mu\sum_{n=1}^{L}(2c_n^{\dagger} c_n-1) +\Delta\sum_{n=1}^{L}\sum_{\ell=1}^{L-1}d_\ell^{-\alpha}\left( c_n c_{n+\ell}+c_{n+\ell}^{\dagger}c_n^{\dagger}\right),
\label{Hlrk}
\ea
was introduced by Vodola {\it et al.}~\cite{vodola14,viyuela16}. Here,  $L$, $w$
$\mu$, and $\Delta$ stand for  the total number of lattice sites, the nearest neighbour hopping amplitude,  the chemical potential, and the strength of the superconducting gap, respectively. We will assume $w$, $\mu$ and $\Delta$ are real. The fermionic annihilation (creation) operators $c_n$ ($c_n^\dagger$) satisfy usual anti-commutation relations $\left\{c_m,c_n\right\}=0$ and $\left\{c_m,c_n^\dagger\right\}=\delta_{mn}$. The superconducting pairing term is a function of the distance $d_\ell$ between any two sites $n$ and $n+\ell$ and decay algebraically characterized by the exponent $\alpha$ ($\alpha>0$). For a closed ring $d_\ell=\text{Min}[\ell, (L-\ell)]$, while for an open chain $d_\ell=\ell$. Although the fermionic number is not conserved but the parity is conserve in the above Hamiltonian
like the corresponding short-range model.

Considering the anti-periodic boundary condition, the above quadratic Hamiltonian can be recast  in the momentum space  with $c_k=({1}/{\sqrt{L}})\sum_n c_n \exp(-ikn) $, where the momentum mode are quantized as $k=({2\pi}/{L})\left(n+1/2\right)$ with $n$ is a positive integer. Then,  the Hamiltonian in Eq.~\eqref{Hlrk} can be written in the following block diagonal form for each momentum mode,
\be
H=\sum_k \left({\begin{array}{cc} c_k^\dagger & c_{-k} \end{array}}\right)H_k
\left(\begin{array}{c} c_k \\ c_{-k}^\dagger \end{array}\right) 
\ee
with 
\be
H_k=\left(\begin{array}{cc} w\cos k+\mu & -i\Delta f_\alpha(k)\\ i\Delta f_\alpha(k) & -w\cos k-\mu \end{array}\right)
\label{ham_k}
\ee
where
\be
f_\alpha(k)=\sum_{\ell}^{L-1}\frac{\sin(k\ell)}{d_\ell^\alpha}.
\label{disf}
\ee
The spectrum of excitations can be obtained through a Bogoliubov transformation,
\be
\left(\begin{array}{c} \eta_k \\ \eta_{-k}^\dagger \end{array}\right)=\left(\begin{array}{cc} \cos \theta_k & i\sin \theta_k\\ i\sin \theta_k & \cos \theta_k \end{array}\right)\left(\begin{array}{c} c_k \\ c_{-k}^\dagger \end{array}\right)
\ee
where $\theta_k$ is given by
\be
\tan (2\theta_k)=-\frac{\Delta f_\alpha(k)}{w\cos k +\mu}.
\label{theta_k}
\ee
In the Bogoluibov basis, the Hamiltonian can be put in the diagonalised form:
\be
H=\sum_k \lambda_\alpha(k)\left(\eta_k^\dagger\eta_k-\frac{1}{2}\right)
\ee
where $\lambda_\alpha(k)$ is the quasiparticle dispersion relation is given by
\be
\lambda_\alpha(k)=\sqrt{(w\cos k +\mu)^2+(\Delta f_\alpha(k))^2}.
\label{lrspectrum}
\ee 
As the Hamiltonian does not commute with the fermionic number, the ground state in momentum basis is given by the vacuum of the Bogoliubov fermions and it has a BCS like form;
\be
\ket{\psi_k}=\prod_k \left(\cos \theta_k-i\sin\theta_k c_k^\dagger c_{-k}^\dagger\right)\ket{0}
\label{gsbcs}
\ee  
where $\ket{0}$ is vacuum of the $c_k$ fermions. In the thermodynamic limit $L\rightarrow\infty$, the momenta assume continuous values in the Brillouin zone and the particle-hole symmetric dispersion relation will be
\be
\lambda_\alpha^\infty(k)=\pm\sqrt{(w\cos k +\mu)^2+(\Delta f_\alpha^\infty(k))^2}.
\label{clrspectrum}
\ee 
where
\ba f_\alpha^\infty(k)&=&-\frac{i}{2}\sum_{\ell=1}^{\infty}\frac{\exp(ik\ell)-\exp(-ik\ell)}{\ell^\alpha} \non \\ &=&\frac{i}{2}\left[\text{Li}_\alpha(e^{-ik})-\text{Li}_\alpha(e^{ik})\right]
\ea
with $\text{Li}_\alpha(x)$ being the poly-logarithmic functions of $x$. (For a thorough discussion on these function, refer to \ct{vodola_th}.)

One can further generalise Hamiltonian in \eqref{Hlrk} to the form: 
\ba
H_{LRK}=-w\sum_{n=1}^{L} d_\ell^{-\beta} \left(c_n^{\dagger}c_{n+1}+c_{n+1}^{\dagger}c_n\right)-\mu\sum_{n=1}^{L}(2c_n^{\dagger} c_n-1) +\Delta\sum_{n=1}^{L}\sum_{\ell=1}^{L-1}d_\ell^{-\alpha}\left( c_n c_{n+\ell}+c_{n+\ell}^{\dagger}c_n^{\dagger}\right),
\label{Hlrkhop}
\ea
with an additional long-range hopping term with exponent $\beta >0$.
Then the Hamiltonian in Eq.~\eqref{ham_k} and the dispersion relation in Eq.~\eqref{lrspectrum} will have similar form with  $\cos k$ getting substituted  by $g_\beta(k)=\sum_{\ell}d_\ell^{-\beta}\cos(k\ell)$. It has, however,
been  show that  the inclusion of  the long-range hopping does not produce any new topological phases rather it may enlarge the topologically non-trivial phase. The effects of long-range hopping~\cite{degottardi13_1} and as well as long-range pairing has been discussed in the Ref.~\cite{alecce17}. 

\subsection{Topological invariant: the winding number }
The above Hamiltonian is time reversal, particle-hole and chiral symmetric and lies in the BDI class of the tenfold way of classification of topological insulators and superconductors \ct{altland97}. The Hamiltonian $H_k$ in Eq.~\eqref{ham_k} can be map to a winding vector $\vec{n}_k=(0,\Delta f_\alpha^\infty(k),w\cos k +\mu)$ in the $y$-$z$ plane by rewriting the Hamiltonian as $H_k=\vec{n_k}\cdot\vec{\sigma}$, where $\vec{\sigma}=\left(\sigma_x,\sigma_y,\sigma_z\right)$ are the Pauli matrices.  Considering the angle $\phi_k=\tan^{-1}\left[\Delta f_\alpha^\infty(k)/(w\cos k +\mu)\right]$ made by the vector $\hat{n}_k$ with the $z$ axis, we can define a winding number as
\be
\nu~=~\frac{1}{2\pi}\oint d\phi_k~=~\frac{1}{2\pi}\int_{0}^{2\pi}dk \left(\frac{d\phi_k}{dk}\right),
\label{wn}
\ee
which is the angle subtended by the vector $\vec{n_k}$, as we go around the Brillouin zone, divided by $2\pi$. The winding number $\nu$ remains unchanged under any adiabatic deformation of the Hamiltonian $H_k$ via continuous changing a parameter of it, while keeping the symmetries of the system intact, without closing the bulk gap. It can only change if we close the gap and reopen it. Therefore, in a gapped phase, it defines a $Z$ valued topological invariant and generally takes integer value. We will call a phase topologically non-trivial if $\nu\neq0$ and non-topological or trivial if $\nu=0$. 

We note that one can also choose the Berry phase \ct{berry84} in 1D or the Zak phase \ct{zak89} to define a topological invariant which has one to one correspondence with the winding number. After making the Hamiltonian in Eq.~\eqref{ham_k} off-diagonal by a unitary transformation, the eigenstate of the lower band $\lvert u_k^-\rangle$ of the system picks up a Berry Phase;
\be
\Phi_z= i\int_{k_0}^{k_0+G} \langle u_k^-\rvert \partial_k\lvert u_k^-\rangle dk
\label{zak}
\ee   
as we considered a adiabatic transport from a particular momentum $k_0$ up to a reciprocal lattice vector $G$ i.e. $k_0+G$. Generally, it is quantized in integer multiple of $\pi$ that characterizes distinct topological phases.

\section{Short-range Kitaev chain}\label{srk}
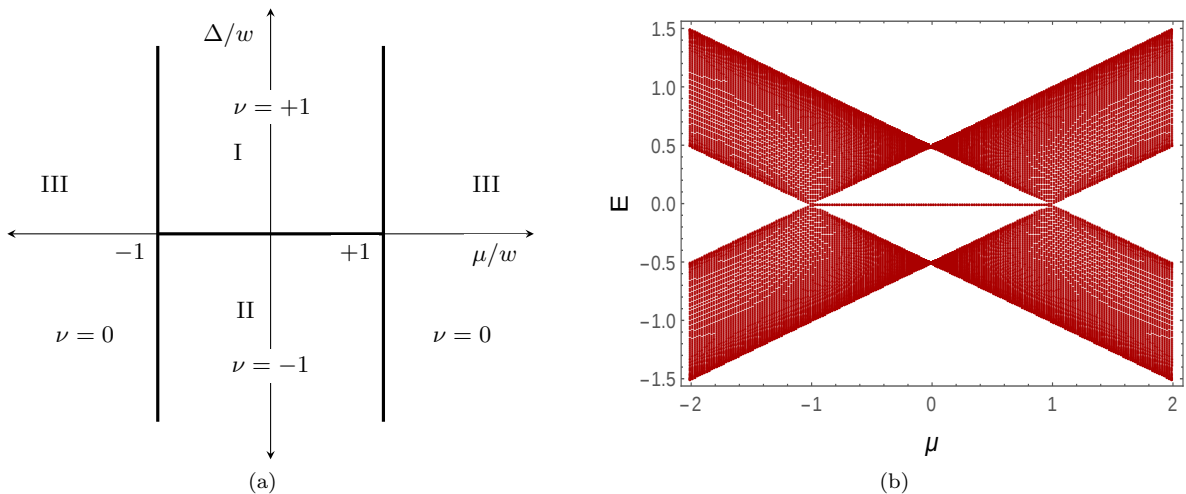
\begin{figure}[]
	\centering 
	\subfigure[]{
	\tikzstyle{arrow} = [ <->,>=stealth]
	\tikzstyle{line} = [ very thick,-,>=stealth]
	\begin{tikzpicture}   
	\draw[arrow] (-3.5,0) node[below=38pt,right=15pt,fill=white] {$\nu=0$} -- (3.5,0) node[below=8pt,left=3pt,fill=white] {$\mu/w$};
	\draw[arrow] (0,-3) -- (0,3) node[below=10pt,left=3pt,fill=white] {$\Delta/w$};  
	\draw[line](-1.5,-2.5) node[above=90pt,left=30pt,fill=white] {III} -- node[above=48pt,right=25pt,fill=white] {$\nu=+1$} (-1.5,2.5) node[below=40pt,right=25pt,fill=white] {I};
	\draw[line](+1.5,-2.5) node[above=90pt,right=30pt,fill=white] {III} -- node[below=38pt,right=15pt,fill=white] {$\nu=0$} (+1.5,2.5) node[below=100pt,left=45pt,fill=white] {II};   
	\draw[line] (-1.5,0) node[below=7pt,left=1pt,fill=white] {$-1$} -- node[below=43pt,fill=white] {$\nu=-1$} (1.5,0)node[below=7pt,left=1pt,fill=white] {$+1$};  
	\end{tikzpicture}
 \label{srpd}}
\hspace{0.5cm}
	\subfigure[]{
		\includegraphics[width=0.43\columnwidth,height=0.33\columnwidth]{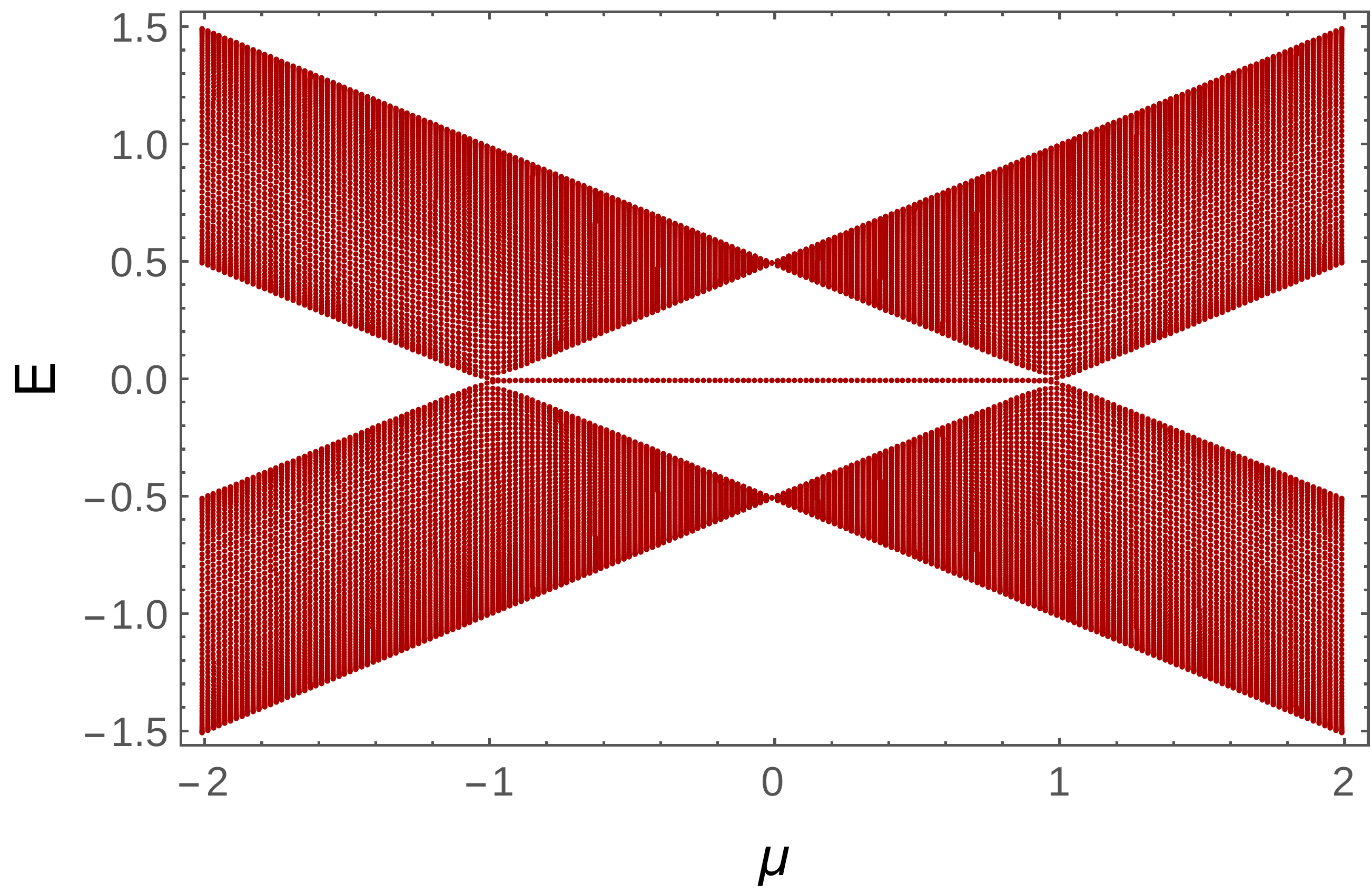}
		\label{srk_open}}
	\caption{(a) Phase diagram of the short-range Kitaev chain  in the $\mu$-$\Delta$ plane with  black thick lines representing  critical lines. Phases I and II are the topologically non-trivial phase whereas phase III is the  non-topological phase. (b) The energy spectrum of an open chain of SRK Hamiltonian in the Majorana basis by diagonalizing the Hamiltonian in Eq.~\eqref{srmajorana} with system size $L=70$, $w=1$ and $\Delta=1$. Note that zero energy Majorana modes are  present in the region $-1<\mu<+1$, i.e., in the topological phase.}
\end{figure}

We  shall now consider  the limit $\alpha\rightarrow \infty$  so that  the pairing term is  limited to nearest neighbours only \cite{kitaev01}. In this limit, the Hamiltonian in Eq.~\eqref{Hlrk} assumes the
simplified form
\ba
H_{SRK}= -w \sum_{n=1}^{L-1}\left(c_n^{\dagger}c_{n+1}+c_{n+1}^{\dagger}c_n\right)-\mu\sum_{n=1}^{L}(2c_n^\dagger c_n-1)+\Delta\sum_{n=1}^{L-1}\left( c_nc_{n+1}+c_{n+1}^{\dagger}c_n^{\dagger}\right),
\label{Hsrk}
\ea
and the corresponding  dispersion relation given in Eq.~\eqref{clrspectrum} reduces to
\be
\lambda(k)=\pm\sqrt{(w\cos k +\mu)^2+(\Delta \sin k)^2}.
\label{srspectrum}
\ee 
It is evident from \eqref{srspectrum},  the spectrum become gap-less at $\mu=\mp w$ for the momentum mode $k=0$ and $k=\pi$, respectively. Therefore, the phase diagram is gapped  everywhere  in the $\mu-\Delta$ plane  except at the critical lines $\mu=\mp w$; see  Fig.~\ref{srpd}. One can further topologically  characterize the phase diagram through  the winding number $\nu$ defined in Eq.~\eqref{wn}. It can be shown
that  there are three topological distinct phases separated by the critical lines $\mu=\pm w$, $\Delta=0$ for SRK model: a trivial phase for $\left|{\mu}/{w}\right|>1$ with $\nu=0$ (Phase III in Fig.~\ref{srpd}) and two topological phases for $\left|{\mu}/{w}\right|<1$ with $\nu=-1$ ( $\nu=1$) when $\Delta/\omega<0$ ($\Delta/\omega>0$)  (Phase I and II in Fig.~\ref{srpd}, respectively).   The phase diagram is expected to be identical for both positive and negative $\mu$; this is  ensured by the fact that  under the unitary transformation $c_n\rightarrow \left(-1\right)^n c_n^{\dagger}$, $\mu$ in  the Hamiltonian in Eq.~\eqref{Hsrk}  goes
to $-\mu$.

One can further   decompose the Dirac fermions ($c_n$) in terms of two  (real) Majorana fermions ($a_n$ and $b_n$) such that  $c_n=\left(a_n-ib_n\right)/2$;  the Majorana operators satisfy the relations $\left\{a_n,a_m\right\}=\left\{b_n,b_m\right\}=2\delta_{mn}$, $\left\{a_n,b_m\right\}=0$ and $a_n^2=b_n^2=1$. Then, the topological trivial and non-trivial phases can also be characterised by the presence and absence of the Majorana zero mode at the boundaries of an open SRK chain. This is the so-called "bulk-boundary" correspondence.

\begin{figure*}[]
	\centering
	\tikzstyle{circle1} = [circle, rounded corners,text centered,minimum width=0.6cm, minimum height=0.6cm, draw=black, fill=red!30,node distance=1.0cm]
	
	\tikzstyle{circle2} = [circle, rounded corners,text centered,minimum width=0.6cm, minimum height=0.6cm, draw=black, fill=blue!30,node distance=1.0cm]

	\tikzstyle{arrow} = [ ultra thick,-,>=stealth]

	\subfigure[]{
		\begin{tikzpicture}
		\node[circle1](a1){\scriptsize{$a_1$}};
		\node[circle2,below of=a1](b1){\scriptsize{$b_1$}};
		\node[circle1,right of=a1](a2){\scriptsize{$a_2$}};
		\node[circle2,below of=a2](b2){\scriptsize{$b_2$}};
		\node[circle1,right of=a2](a3){\scriptsize{$a_3$}};
		\node[circle2,below of=a3](b3){\scriptsize{$b_3$}};
		\node[circle1,right of=a3](a4){\scriptsize{$a_4$}};
		\node[circle2,below of=a4](b4){\scriptsize{$b_4$}};
		\node[circle1,right of=a4](a5){\scriptsize{$a_5$}};
		\node[circle2,below of=a5](b5){\scriptsize{$b_5$}};
		\node[circle1,right of=a5](a6){\scriptsize{$a_6$}};
		\node[circle2,below of=a6](b6){\scriptsize{$b_6$}};
		\node[circle1,right of=a6](a7){\scriptsize{$a_7$}};
		\node[circle2,below of=a7](b7){\scriptsize{$b_7$}};
		\node[circle1,right of=a7](a8){\scriptsize{$a_8$}};
		\node[circle2,below of=a8](b8){\scriptsize{$b_8$}};
		\node[circle1,right of=a8](a9){\scriptsize{$a_9$}};
		\node[circle2,below of=a9](b9){\scriptsize{$b_9$}};
		\node[circle1,right of=a9](a10){\scriptsize{$a_{10}$}};
		\node[circle2,below of=a10](b10){\scriptsize{$b_{10}$}};
		\draw [arrow] (a2) -- node[below,left] {$J_x$} (b1);
		\draw [arrow] (a3) -- node[below,left] {$J_x$}(b2);
		\draw [arrow] (a4) -- (b3);
		\draw [arrow] (a5) -- (b4);
		\draw [arrow] (a6) -- (b5);
		\draw [arrow] (a7) -- (b6);
		\draw [arrow] (a8) -- (b7);
		\draw [arrow] (a9) -- (b8);
		\draw [arrow] (a10) -- (b9);
		\end{tikzpicture} \label{mzma}
	}
	
	\subfigure[]{
		\begin{tikzpicture}
		\node[circle1](a1){\scriptsize{$a_1$}};
		\node[circle2,below of=a1](b1){\scriptsize{$b_1$}};
		\node[circle1,right of=a1](a2){\scriptsize{$a_2$}};
		\node[circle2,below of=a2](b2){\scriptsize{$b_2$}};
		\node[circle1,right of=a2](a3){\scriptsize{$a_3$}};
		\node[circle2,below of=a3](b3){\scriptsize{$b_3$}};
		\node[circle1,right of=a3](a4){\scriptsize{$a_4$}};
		\node[circle2,below of=a4](b4){\scriptsize{$b_4$}};
		\node[circle1,right of=a4](a5){\scriptsize{$a_5$}};
		\node[circle2,below of=a5](b5){\scriptsize{$b_5$}};
		\node[circle1,right of=a5](a6){\scriptsize{$a_6$}};
		\node[circle2,below of=a6](b6){\scriptsize{$b_6$}};
		\node[circle1,right of=a6](a7){\scriptsize{$a_7$}};
		\node[circle2,below of=a7](b7){\scriptsize{$b_7$}};
		\node[circle1,right of=a7](a8){\scriptsize{$a_8$}};
		\node[circle2,below of=a8](b8){\scriptsize{$b_8$}};
		\node[circle1,right of=a8](a9){\scriptsize{$a_9$}};
		\node[circle2,below of=a9](b9){\scriptsize{$b_9$}};
		\node[circle1,right of=a9](a10){\scriptsize{$a_{10}$}};
		\node[circle2,below of=a10](b10){\scriptsize{$b_{10}$}};
		\draw [arrow] (a1) -- node[below=0pt,right=1pt] {$J_y$}(b2);
		\draw [arrow] (a2) -- node[below=0pt,right=1pt] {$J_y$}(b3);
		\draw [arrow] (a3) -- (b4);
		\draw [arrow] (a4) -- (b5);
		\draw [arrow] (a5) -- (b6);
		\draw [arrow] (a6) -- (b7);
		\draw [arrow] (a7) -- (b8);
		\draw [arrow] (a8) -- (b9);
		\draw [arrow] (a9) -- (b10);
		\end{tikzpicture}\label{mzmb}
	}
	
	\subfigure[]{
		\begin{tikzpicture}
		\node[circle1](a1){\scriptsize{$a_1$}};
		\node[circle2,below of=a1](b1){\scriptsize{$b_1$}};
		\node[circle1,right of=a1](a2){\scriptsize{$a_2$}};
		\node[circle2,below of=a2](b2){\scriptsize{$b_2$}};
		\node[circle1,right of=a2](a3){\scriptsize{$a_3$}};
		\node[circle2,below of=a3](b3){\scriptsize{$b_3$}};
		\node[circle1,right of=a3](a4){\scriptsize{$a_4$}};
		\node[circle2,below of=a4](b4){\scriptsize{$b_4$}};
		\node[circle1,right of=a4](a5){\scriptsize{$a_5$}};
		\node[circle2,below of=a5](b5){\scriptsize{$b_5$}};
		\node[circle1,right of=a5](a6){\scriptsize{$a_6$}};
		\node[circle2,below of=a6](b6){\scriptsize{$b_6$}};
		\node[circle1,right of=a6](a7){\scriptsize{$a_7$}};
		\node[circle2,below of=a7](b7){\scriptsize{$b_7$}};
		\node[circle1,right of=a7](a8){\scriptsize{$a_8$}};
		\node[circle2,below of=a8](b8){\scriptsize{$b_8$}};
		\node[circle1,right of=a8](a9){\scriptsize{$a_9$}};
		\node[circle2,below of=a9](b9){\scriptsize{$b_9$}};
		\node[circle1,right of=a9](a10){\scriptsize{$a_{10}$}};
		\node[circle2,below of=a10](b10){\scriptsize{$b_{10}$}};
		\draw [arrow] (a1) -- node[below=0pt,right=1pt] {$\mu$}(b1);
		\draw [arrow] (a2) -- node[below=0pt,right=1pt] {$\mu$}(b2);
		\draw [arrow] (a3) -- (b3);
		\draw [arrow] (a4) -- (b4);
		\draw [arrow] (a5) -- (b5);
		\draw [arrow] (a6) -- (b6);
		\draw [arrow] (a7) -- (b7);
		\draw [arrow] (a8) -- (b8);
		\draw [arrow] (a9) -- (b9);
		\draw [arrow] (a10) -- (b10);
		\end{tikzpicture} \label{nomzm}
	}	
	
	\caption{This figure is showing a schematic diagram of the presence and absence of Majorana zero modes in different extreme limits for an open chain of SRK model with the total number of sites $L=10$ and different parameter values (a) $J_x=(\Delta+w)/2 >0$, $J_y=0$, $\mu=0$ (b) $J_y=(\Delta-w)/2 > 0$, $J_x=0$, $\mu=0$ and (c) $J_x=J_y=0$, $\mu\ne0$.   } 
	\label{fig:lnlblock} 
\end{figure*}
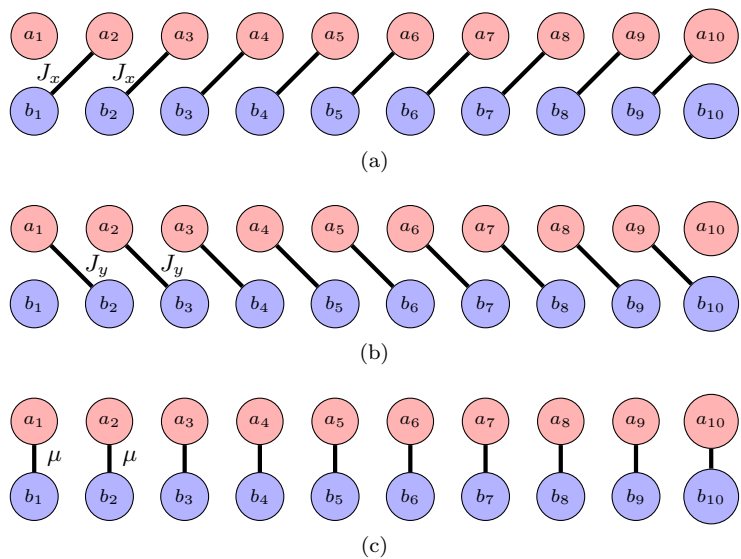

In terms of Majorana fermions one can rewrite the Hamiltonian in Eq.~\eqref{Hsrk} with an open boundary condition as,
\ba
H_{SRK}=i\sum_{n=1}^{L-1}\left[-J_xb_na_{n+1}-J_y a_nb_{n+1}\right]+i\mu\sum_{n=1}^{L}a_nb_n,
\label{srmajorana}
\ea
with $J_x=\frac{1}{2}\left(\Delta+w\right)$ and $J_y=\frac{1}{2}\left(\Delta-w\right)$.
In the limit of extreme weak pairing; $\mu=0$, $J_y=0$ and $J_x>0$ ($\Delta=w>0$), the tight binding Hamiltonian in Eq.~\eqref{srmajorana} connects the inter site Majorana fermions in a staggered fashion as shown in the Fig.~\ref{mzma}. This yields two Majorana fermions $a_1$ and $b_L$ are isolated at the left and right end of an open chain, respectively, not appearing in the Hamiltonian, and thus, one has  two Majorana zero modes (MZMs) at the two ends of the chain. On the other hand, in the other limit of  $J_x=0$ and $J_y>0$ ($\Delta=-w>0$) with $\mu=0$, there exist   two MZMs $b_1$ and $a_L$ at the left and right ends, respectively, (as shown in the Fig.~\ref{mzmb}). The presence of MZMs at either ends in these two limits refers the topological non-trivial phase I and II in Fig.~\ref{srpd} with winding number $\nu=\pm1$, respectively.  Finally, in the extreme strong pairing limit between intra-site Majoranas; $\mu\neq0$ and $J_x=J_y=0$ ($\Delta=w=0$), the Majorana fermions $a_n$ and $b_n$ are pairwise connected and there are no MZMs at either end of an open chain (see Fig.~\ref{nomzm}), yields a topological trivial phase with $\nu=0$ (phase III in Fig.~\ref{srpd}). This topological behavior is protected by the gap in the spectrum and does not alter as long as a gap-less quantum critical point is crossed \ct{degottardi13}.

The ground state of the topological non-trivial phases is two fold degenerate, which is associated with the isolated MZMs. The two isolated MZMs can together be written in terms of an ordinary fermion $f=\left(a+ib\right)/\sqrt{2}$ (for which $f \neq f^\dagger$); the corresponding zero energy state may be occupied or unoccupied ($f^\dagger f = 1$ or $0$). This gives rise to two possible ground states which are degenerate for an infinitely long chain. Hence the ground state of the SRK  chain is two-fold degenerate, but the two states have different parities depending on whether the zero energy state is occupied or not. Thus, choosing one of the two degenerate states amounts to breaking the $\mathbb{Z}_2$ symmetry of the ground state. 

Using the Jordan-Wigner transformation \cite{lieb64}, one can also map the fermionic system in Eq.~\eqref{srmajorana} in to spin-$1/2$ XY chain in a transverse magnetic field,  Introducing the Jordan-Wigner transformations from the spin-$1/2$ operators to Majorana operators:
\ba
a_n=\left(\prod_{j=1}^{n-1} \sigma_j^z\right) \sigma_n^x, ~~~\text{and}~~~
b_n=\left(\prod_{j=1}^{n-1} \sigma_j^z\right) \sigma_n^y
\ea
where $\sigma_n^a$ ($a=x/y/z$) denote the Pauli matrices at site $n$, then the Hamiltonian in Eq.~\eqref{srmajorana} can be written as
\be
H=-\sum_{n=}^{L-1} \left( J_x \sigma_n^x \sigma_{n+1}^x + J_y \sigma_n^y \sigma_{n+1}^y \right) -\mu\sum_{n=1}^{L}\sigma_n^z.
\ee
Referring to Fig.\ref{srpd},  the phases I and II are $\mathbb{Z}_2$ symmetry-broken ferromagnetic phases, while the phase III is paramagnetic.

\section{Long-range limit, correlations and phase diagram}
\label{sec_topology}
It is evident that for any finite value of the exponent $\alpha$, the Hamiltonian in Eq.~\eqref{Hlrk} can not be mapped to XY model by the Jordan Wigner transformation. Furthermore, the transformation $c_n\rightarrow\left(-1\right)^n c_n^\dagger$ no longer connects the negative $\mu$ to positive $\mu$, thus the phase diagram is also not symmetric around $\mu=0$ for any finite $\alpha$ unlike the SRK model.

\subsection{Critical lines and phase diagram}
\label{sec_critical}
In order to determine the phase diagram in the limit $L \to \infty$, we recall the dispersion relation:
\be
\lambda_\alpha^\infty(k)=\pm\sqrt{(\cos k +\mu)^2+(f_\alpha^\infty(k))^2},
\label{spectrum}
\ee  
where  we have chosen $w=\Delta=1/2$, without loss of generality. The critical lines can be obtained by studying the zeros and divergences of the function
\be
f_{\alpha}^{\infty}(k)=\frac{i}{2}\left[\text{Li}_\alpha(e^{-ik})-\text{Li}_\alpha(e^{ik})\right],
\ee
For $\alpha>1$, the function $f_{\alpha}^{\infty}(k)=0$ for $k=0$ and $k=\pi$. Thus, the dispersion goes to zero for these two critical modes $k=0$ and $k=\pi$ at both the critical lines $\mu=\mp1$, respectively. On
the contrary, the situation changes drastically when $\alpha < 1$, the function $f_{\alpha}^{\infty}(k)$ is only zero for $k=\pi$, yields the line $\mu=+1$ is gapless , whereas the line $\mu=-1$ is gapped. This is because of  the fact that the function $f_{\alpha}^{\infty}(k)$ diverges as $k\rightarrow0$ for $\alpha<1$  and hence $\lambda_\alpha^\infty \sim k^{-(1-\alpha)}$ in the limit $k\rightarrow0$. Consequently,  the energy of the system also becomes super-extensive for $\alpha<1$. Therefore, in the case of LRK model, there are {only}  {\it two} different sectors in the phase diagram separated by the line $\alpha=1$.

\subsection{Correlation function and Entanglement entropy}\label{cor_ee}
The scaling of the correlation functions and the entanglement entropy in the ground state  of the LRK chain  has been discussed by Vodola $et~ al.$,  in \cite{vodola14}. Let us consider the one body two point correlation function $C_r=\langle\psi_0\rvert c_r^\dagger c_0\lvert\psi_0\rangle$ at a distance $r$, where the expectation value is taken over the ground state $\ket{\psi_0}$ in the fermionic basis;  this can be exactly found using the ground state defined in Eq.~\eqref{gsbcs} as follows,
\be
C_r~=~\frac{1}{L}\sum_k e^{-ikr} \langle \psi_k\rvert c_k^\dagger c_k\lvert \psi_k\rangle~=~\frac{1}{L}\sum_k e^{-ikr}\sin^2\theta_k
\ee
where $\theta_k$ is given in Eq.~\eqref{theta_k}. Similarly, we can also calculate the anomalous two point correlation function $F_r=\langle \psi_0\rvert c_r^\dagger c_0^\dagger\lvert\psi_0\rangle$, which is given by,
\be
F_r~=~\frac{1}{L}\sum_k e^{-ikr}i\sin\theta_k\cos\theta_k
\ee
These correlators can be solved analytically as a function of $\mu$ and $\alpha$ in the thermodynamic limit as rigorously shown in Ref.~\cite{vodola14}.. Interestingly, a hybrid exponential-algebraic decay of the two point correlators has been found. More specifically, it is exponential in the short distance and have an algebraic tail in the large distance for $\alpha>1$ and the exponential decay dominates for as $\alpha$
increases; on the other hand, it has purely an algebraic decay for $\alpha<1$ i.e., when system become sufficiently long-ranged. This algebraic decays of the correlation function is unexpected, because  the system is  in a gapped (non-critical) phase. This kind of algebraic decays was also observed in long-range Ising models~~\cite{koffel12,hauke10,muller12}. As the LRK model is free fermionic, all the higher order correlators, such as density-density correlations etc., can be obtained using the above two point correlators using Wick's theorem.

We shall now proceed to discuss the entanglement entropy (EE) between two parts of the LRK chain  as a measure of the entanglement present in the system. Let us divided the system into two parts: $A$ with system size $\ell$ and $B$ with system size $L-\ell$. The reduce density matrix of the subsystem $A$ can be calculated by tracing out all the degrees of freedom corresponds to the rest of the system $B$; 
\be
\rho_{\ell}~=~\Tr_B\rho 
\ee
where $\rho=\ket{\psi_0}\bra{\psi_0}$ is the density matrix for the whole system. Then, the entanglement entropy $S_\ell$ can be obtained by calculating the Von-Neumann entropy of the reduced density matrix $\rho_\ell$ as
\be
S_\ell~=~-\Tr \left(\rho_\ell\log\rho_\ell\right).
\ee 
For a free fermionic model, the EE can also be solved using the two point correlators $C_{mn}=\bra{\psi_0} c_m^\dagger c_n\ket{\psi_0}$ and $F_{mn}=\bra{\psi_0} c_m^\dagger c_n^\dagger\ket{\psi_0}$, where $m,n=1,2,\dots\ell$ by constructing a $2\ell\times2\ell$ correlation matrix~ $\mathcal{C}$ as 
\be
\mathcal{C}~=~\left(\begin{array}{cc} I-C & F\\ F^{\dagger} & C \end{array}\right).
\ee  
where $I$ is the $\ell\times\ell$ identity matrix. Then the EE is given by $S_\ell=-\sum_{i=1}^{2\ell}\lambda_i\log\lambda_i$, where $\lambda_i$ are the eigenvalues of the correlation matrix matrix $\mathcal{C}$ \ct{peschel03}.

In the ground state of a gapped Hamiltonian having short-range interactions, the EE generally obeys an area law~\cite{eisert10} due to finite correlation length ($\xi$) present in the system, 
\be
S_\ell\sim\frac{c}{3}\log\xi,
\ee  
where $c$ is the central charge. As a result, for one dimensional short-range systems, the EE is given by to a non-universal constant value which is independent of the subsystem size $\ell$. On the other hand, at the criticality, the EE shows a universal logarithmic divergence with the subsystem size~\cite{holzhey94,vidal03,latorre04} as,
\be
S_\ell\sim\frac{c}{3}\log\ell.
\label{log_ee}
\ee

\begin{figure}[]
	\centering
	\subfigure[]{
		\includegraphics[width=0.45\columnwidth,height=0.35\columnwidth]{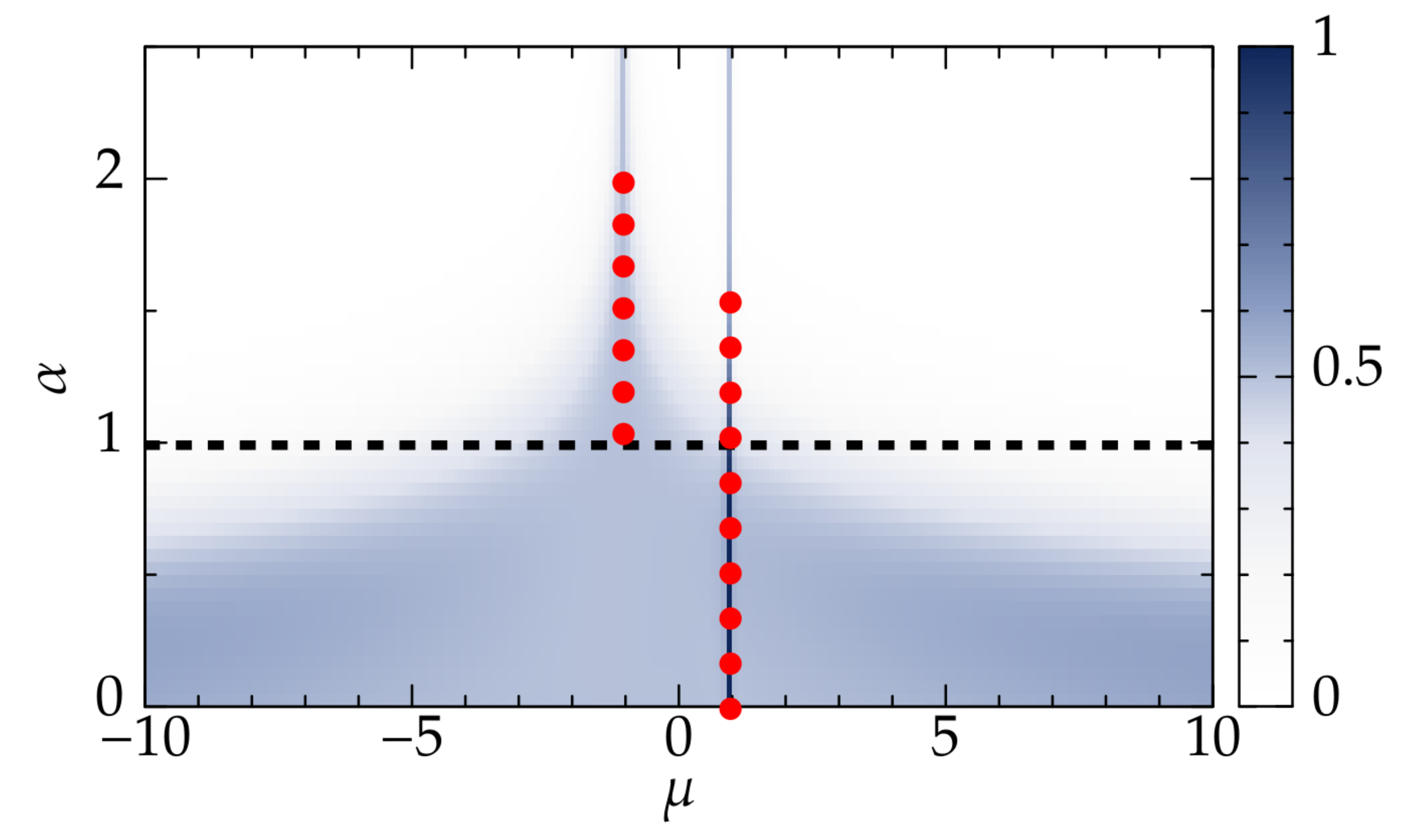}
		\label{ceff_diagram}}
	\hspace{0.05cm}
	\subfigure[]{
		\includegraphics[width=0.45\columnwidth,height=0.36\columnwidth]{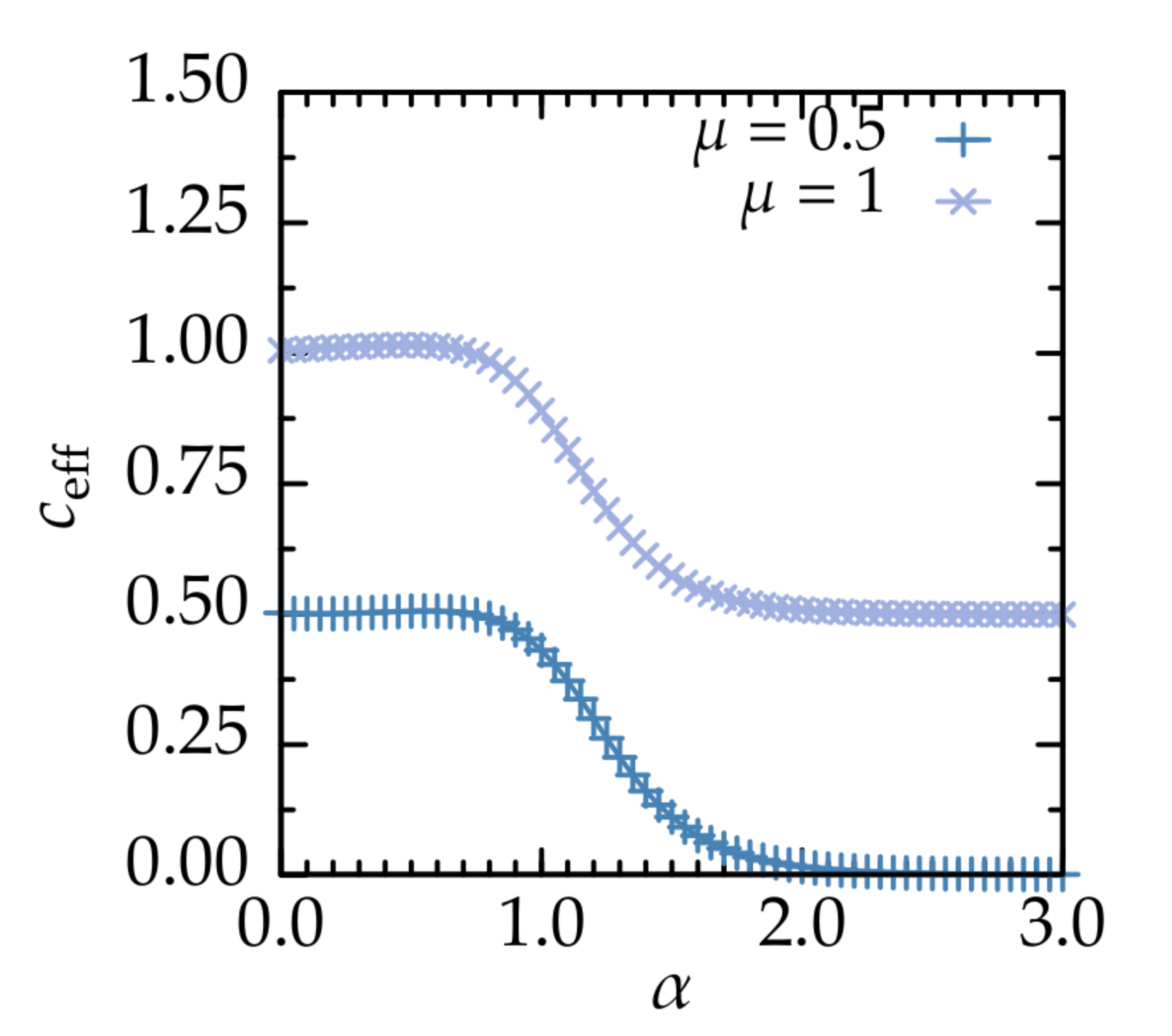}
		\label{ceff_alpha}}
	\caption{(a) The figure shows the phase diagram  in the $\mu-\alpha$ plane obtained from the effective central charge analysis fitting the von Neumann entropy $S_{L/2}$ using Eq.~\ref{cft_cc_effective}.   There exist two gap-less conformal field theories with $c=1/2$ appear for $\mu=1$ (for $\alpha >3/2$)and $\mu =-1$ (for $\alpha >2$).The broken conformal symmetry gap-less lines  are shown by the red dots. Horizontal dashed line separates two regions: for $ \alpha  > 1$, the  correlation functions display a hybrid exponential-algebraic decay while for  $\alpha  < 1$, the decay is purely of algebraic nature. (b) Variation of effective central charge with $\alpha$ for two different values of $\mu$; along the critical line $\mu=1$ and in a gapped phase $\mu=0.5$. One observes  the violation of the area law of EE (for $\mu=0.5$) and the increase from the central charge of $1/2$ to $1$ (for 
		$\mu=1$). (Figure taken from Ref.~\cite{vodola_th} with permission from the author.)}
	\label{ceff}
\end{figure}

Using the  conformal field theory results,  a universal scaling of the EE at the conformally invariant QCP of a one dimensional short-range model with the periodic boundary condition can be obtained \cite{calabrese04} as:
\be
S_\ell~=~\frac{c}{3}\log_2\left(\frac{L}{\pi}\sin\left(\frac{\pi\ell}{L}\right)\right)+a
\label{cft_ee}
\ee 
where $c$ is the central charge of the theory and $a$ is non-universal constant. The above equation \eqref{cft_ee} reduces to the Eq.~\ref{log_ee} in the limit $\ell<<L$. For the SRK chain, $c=1/2$ along the critical lines $\mu=\pm1$. The above scaling of EE can be used to find the the central charge of non-integrable models by calculating the EE using numerical techniques.

In the Ref~\cite{vodola14}, the EE for a half chain $S_{L/2}$ has been calculated for the LRK model and also to study the violation of the area law every where  in the $\mu$-$\alpha$ plane a finite size scaling 
ansatz has been proposed:

\be
S_\ell~=~\frac{c_{\rm eff}}{3}\log_2\left(\frac{L}{\pi}\sin\left(\frac{\pi\ell}{L}\right)\right)+a,
\label{cft_cc_effective}
\ee
where $c_{\rm eff}$ is the effective central charge. (It should however be noted that the scaling form in Eq.~\eqref{cft_ee} is strictly valid only at the gapless conformally invariant critical point.)

The result, thus obtained, is presented  in Fig.~\ref{ceff} where  the value of $c_{\text{eff}}$ is depicted on the phase diagram of the LRK model. (i) It has been found that, for $\alpha>1$ in the gapped region ($\mu\ne\pm1$) $c_{\text{eff}}=0$ and the EE obeys the area law. (ii) On the other hand, a violation of area law occurs when $\alpha<1$ as $c_{\text{eff}}\ne0$ in the gapped region. This violation of area law of EE in gapped phase is an artefact of the dominating long-range interaction ($\alpha<1$) present in the system. More importantly, a breaking of conformal symmetry has been found along the critical line $\mu=+1$ ($\mu=-1$) for $\alpha<3/2$ ($\alpha<2$) as $c_{\text{eff}}>1/2$ as shown  Fig~\ref{ceff_diagram}. In Fig.~\ref{ceff_alpha}, the variation of $c_{\rm eff}$ with the parameter $\alpha$ is shown for the gap-less line $\mu=+1$ and a gapped phase $\mu=0.5$.
It is further observed that for $\alpha<1$,  $c_{\text{eff}}=1$ along the the critical line $\mu=+1$ and  this indicates that the nature of the phase transition along the critical line $\mu=+1$ changes drastically for the LRK chain in contrast to the SRK chain; in the latter situation $c=1/2$ at both the critical lines $\mu=\pm 1$.  This shows that the nature of transition becomes Dirac like ($c_{\rm eff}=1$) for $\alpha <1$, while
it is Majorana like ($c=1/2$) for the SRK chain. 

We note in the passing that similar scaling ansatz as in Eq.~\eqref{cft_cc_effective} has also been employed to study the quantum phase transition and the associated central charge in a long-range transverse Ising chain with antiferromagnetic interactions \ct{koffel12}. The existence of two gapped phases have been
established: one is  dominated by the transverse field,
having  quasi-long-range order while the other  has a long-range Neel
ordered ground states separated by quantum critical points.  Further,  in the phase with quasi-long-range
order the ground states exhibit exotic corrections to the area law for the entanglement entropy coexisting
with gapped entanglement spectra. In the SR regime, the central charge has been found to exceed $1/2$.

\subsection{Topological phase diagram}
Here, we will focus on the topological aspects of the phase diagram related to the presence and absence of MZMs or massive Dirac modes (MDMs) and compare with that of the SRK chain.\\

\begin{figure}[]
	\centering
	\subfigure[]{
		\includegraphics[width=0.31\columnwidth,height=0.29\columnwidth]{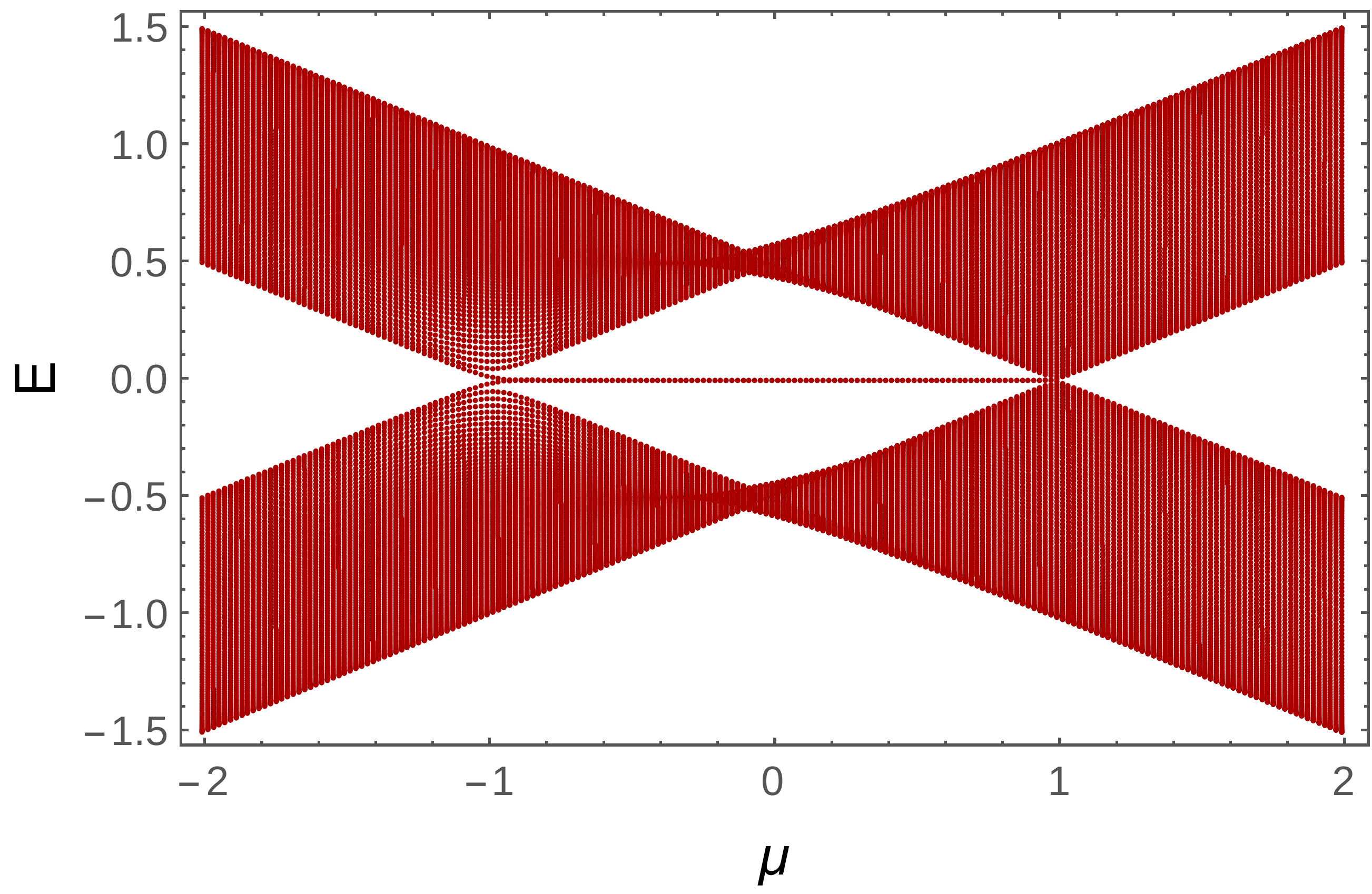}
		\label{mzm}}
	\hspace{0.05cm}
	\subfigure[]{
		\includegraphics[width=0.31\columnwidth,height=0.287\columnwidth]{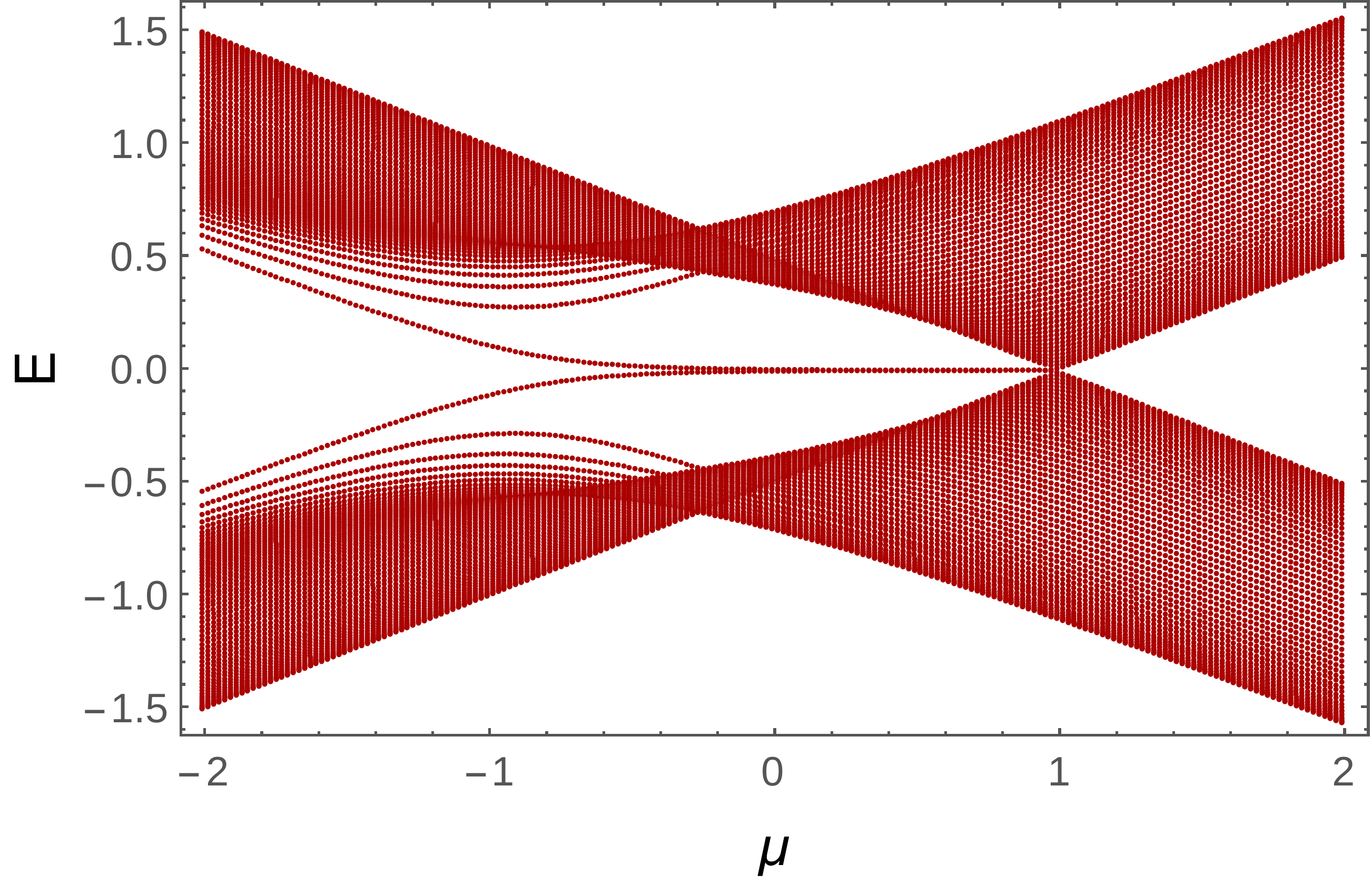}
		\label{crossover1}}
	\hspace{0.05cm}
	\subfigure[]{
		\includegraphics[width=0.31\columnwidth,height=0.288\columnwidth]{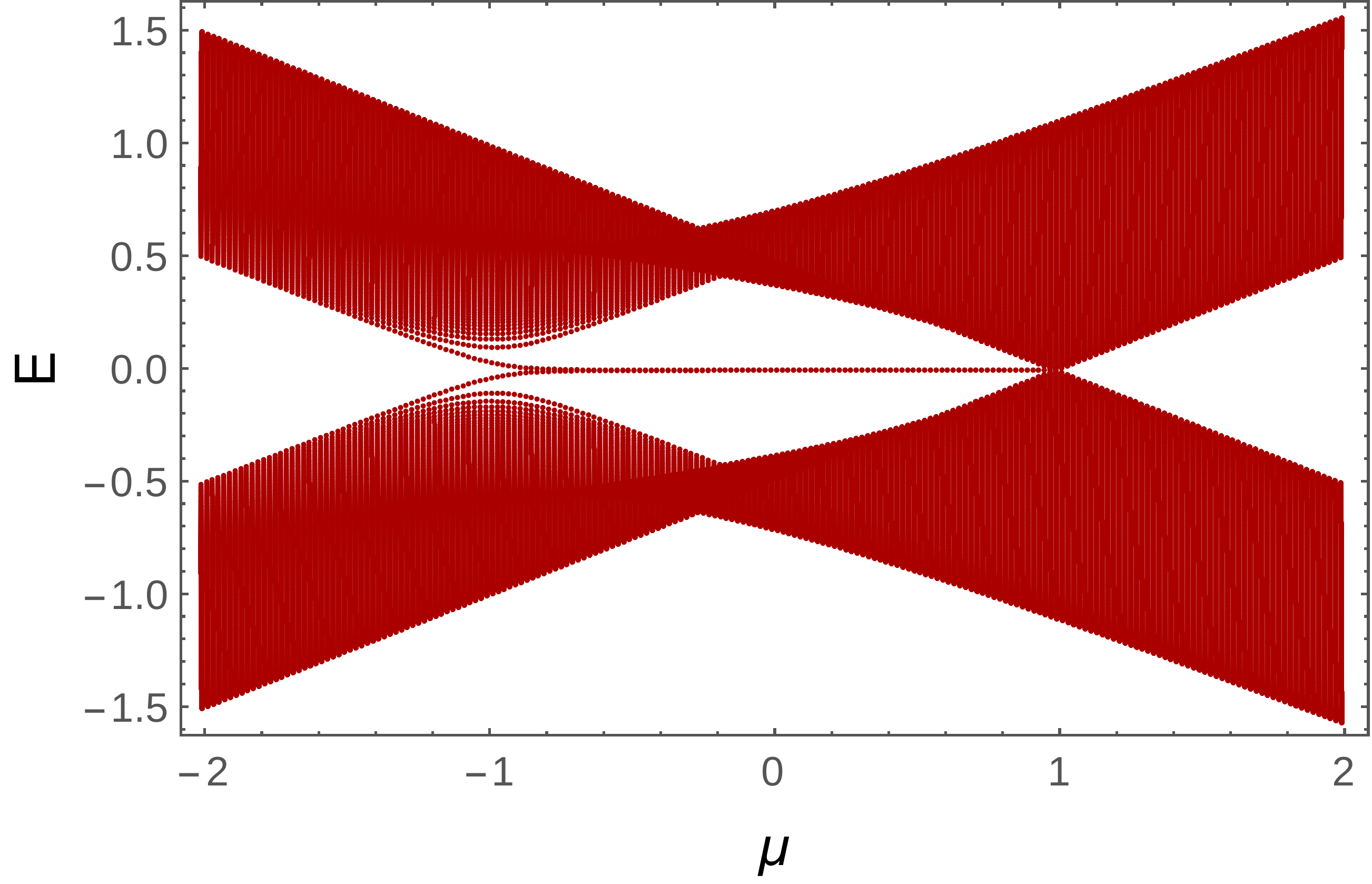}
		\label{crossover2}}
	\begin{picture}(1,0.95238095)%
	\put(-190,155){\color[rgb]{0,0.2,1}\makebox(0,0)[lb]{$ \alpha=2.5, L=100 $}}
	\put(-20,155){\color[rgb]{0,0.2,1}\makebox(0,0)[lb]{$ \alpha=1.4, L=100 $}}
	\put(150,155){\color[rgb]{0,0.2,1}\makebox(0,0)[lb]{$ \alpha=1.4, L=1000 $}}%
	\end{picture}
	\caption{Energy spectrum of the LRK Hamiltonian in Majorana basis for different values of the exponent $\alpha$ and system size $L$ to show that LRK chain is topologically equivalent to SRK chain for $\alpha>1$. (a) For $\alpha=2.5$ and $L=100$,  the spectrum  is similar to  the Fig.~\ref{srk_open} of SRK chain. (b) For $\alpha=1.4$ and  $L=100$, the spectrum looks very different from the energy spectrum of the SRK chain with a crossover region from MZMs to MDMs. But these MDMs do not survive and  the spectrum resembles that of  the of SRK chain as we increase the system size $L$ ($L\rightarrow\infty$), as shown in (c) for  $L=1000$ with the same value of  $\alpha$. (After Ref. \ct{bhattacharya19}).}
	\label{eff_srk}
\end{figure}
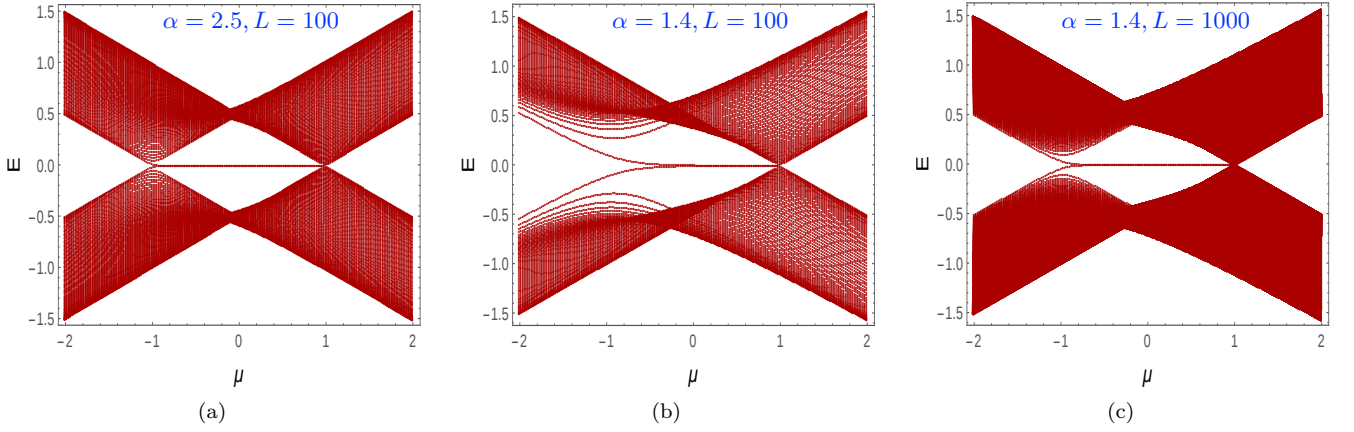

\textbf{(a) Majorana sector ($\alpha >1$):} To probe the topological aspects of the phase diagram, we will divide this region into two sub-regions $\alpha>3/2$ and $1<\alpha<3/2$. This is because of that the group velocity $v_g=\partial_k\lambda(k)$ scales as $v_g\sim k^{-(3-2\alpha)}$ as $k\rightarrow0$ and hence diverges for $\alpha<3/2$ although the dispersion relation $\lambda_\alpha(k)$ is not divergent as long as $\alpha\geq1$.

In the region $\alpha>3/2$, the energy spectrum is topologically equivalent to the SRK chain. As shown in the Fig.~\ref{mzm}, the system hosts MZMs at the two ends of an open chain for $-1<\mu<1$.  The LRK
chain is in  topologically non-trivial  phase with winding number $\nu=1$ when studied with a periodic boundary condition. This is incongruence with the bulk boundary correspondence. Otherwise ($|\mu|>1$), the system is topologically trivial with winding number $\nu=0$ and there is no MZMs present in the open chain (see Appendix~\ref{app_wn}).

On the other hand, for $1<\alpha<3/2$, for small system size $L$,  Fig.~\ref{crossover1} apparently shows a crossover region from MZMs to MDMs  between $\mu=-1$ to $\mu=1$ in   the energy spectrum for
an open chain . But this crossover region does not really exist in the thermodynamic limit $L\rightarrow\infty$ as shown in the Fig~\ref{crossover2} for a larger value of $L=1000$. As the group velocity diverges as $k\rightarrow0$, the spectrum becomes highly dispersive near $k=0$ and it takes very large system size $L$ to realise the correct energy spectrum of the system. Thus for  $1<\alpha<3/2$, the energy spectrum topologically resembles with the SRK chain with only MZMs present in the region $-1<\mu<1$. Therefore, the LRK chain becomes effectively short-range in terms of the presence and absence of MZMs for entire range $\alpha>1$~\cite{lepori171,giuliano18,bhattacharya19}
in contrast to the claim made in \ct{viyuela16}. \\

\begin{figure}[]
	\centering
	\subfigure[]{
		\includegraphics[width=0.35\columnwidth,height=0.3\columnwidth]{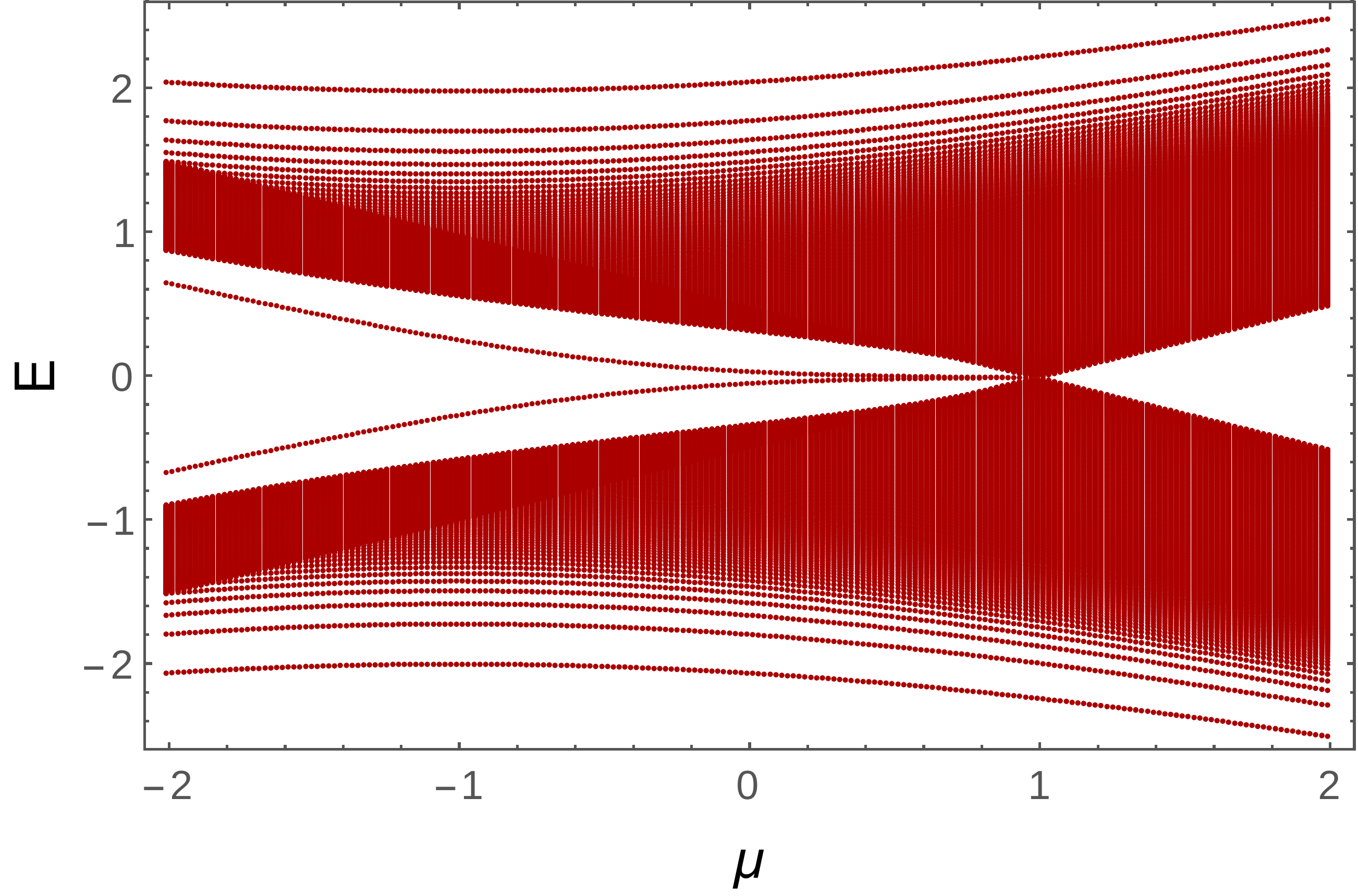}
		\label{mdm}}
	\hspace{0.5cm}
	\subfigure[]{
		\tikzstyle{arrow} = [ ->,>=stealth]
		\tikzstyle{line} = [ very thick,-,>=stealth]
		
		\begin{tikzpicture}   
		\draw[arrow] (-2,-2) --node[above=20pt,right=40pt]{$\nu=-1/2$} (4,-2)node[below=8pt,left=2pt,fill=white] {$\mu$} ; 
		\draw[arrow] (1,-2)node[below=1pt] {$0$} node[above=20pt,left=10pt] {$\nu=+1/2$}  --  (1,2.5)node[below=12pt,left=2pt,fill=white] {$\alpha$};
		\draw [thick,dashed] (-2,-0.75)node[below=0pt,left=0pt,fill=white] {$\alpha=1$} node[above=30pt,right=10pt] {$\nu=0$} -- node[above=30pt,fill=white] {$\nu=\pm1$}(4,-0.75)node[above=30pt,left=10pt] {$\nu=0$};
		\draw[line]node[below=64pt,left=0pt] {$-1$}(-0.25,-0.75)  --  (-0.25,2.5);
		\draw[line](2.25,-2)node[below=1pt] {$+1$}  --  (2.25,2.5);
		\end{tikzpicture}
		
		\begin{picture}(1,0.95238095)%
		\put(-370,142){\color[rgb]{0,0.2,1}\makebox(0,0)[lb]{$ \alpha=0.8, L=1000 $}}
		\end{picture}
		\label{lrpd}}
	\caption{(a) Energy spectrum of the LRK Hamiltonian with an open boundary condition in Majorana basis for $\alpha=0.8$, $L=1000$. Topological massive Dirac modes occurs when $\mu<1$, which is a manifestation  of the long-range pairing in LRK chain. There is no such MDM present for $\mu>1$. (b) Topological phase diagram of the LRK chain  in the $\mu-\alpha$ plane as determined by the
		winding number.  The line $\alpha=1$  marks the boundary between the long-range and the short-range topological character. Note that the critical line 
		at $\mu =-1$ does not exist for $\alpha <1$ as discussed in Sec \ref{sec_critical}. (After Ref. \ct{bhattacharya19})}
\end{figure}
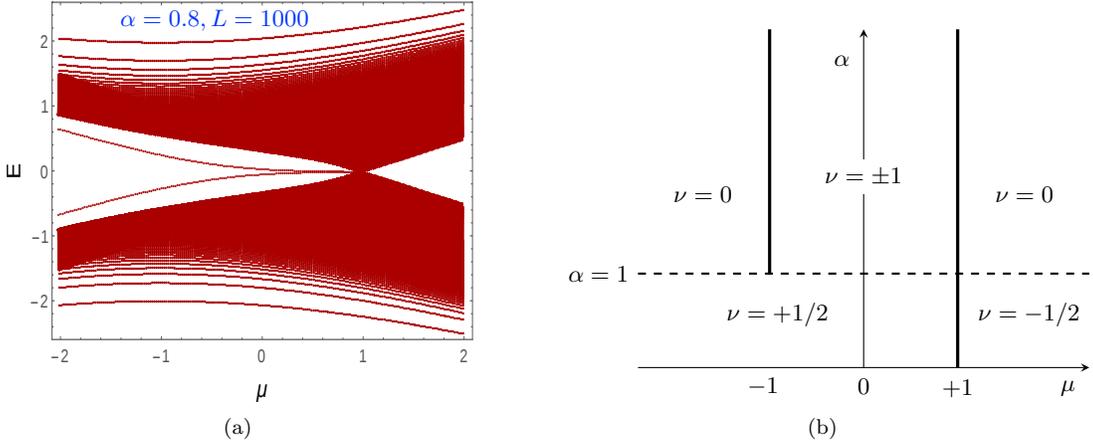

\textbf{(b) Massive Dirac sector:} Let us focus our attention on the region $\al < 1$ when the long-range nature of interaction truly dominates; here, the phase diagram has already shown  to be drastically different from the conventional SRK model. In this region, for all $\mu >+1$ the system with open boundary conditions is in a trivial phase with no end modes, while for all $\mu < +1$, 
this system hosts topological massive Dirac modes (MDMs) at the ends (see Fig.~\ref{mdm}). These MDMs appear due to the coupling induced between the two MZMs at the 
two distant ends due to the presence of long-range pairing, rendering them highly non-local. Moreover, although the Dirac 
mode is massive, it is still topological and 
protected by the bulk gap. This non-local topological quasiparticle is also protected 
by the fermionic parity because the ground state of the system in this phase still retains its even parity; populating the MDM which is the first 
excited state of the system would then require a change in the fermionic parity 
from even to odd.

Since no discrete symmetry has been broken by the inclusion of the long-range pairing, the system still belongs to the BDI symmetry class. However, the winding number $\nu$, evaluated for the periodic
boundary condition, is modified by the topological singularity at $k=0$ generated by the long-range pairing. This happens because at $k=0$ both the energy dispersion $\lambda_\alpha(k)$ and the group velocity $v_g=\partial_k\lambda_\alpha(k)$ diverge as the integrand in the definition of the Berry/Zak phase in Eq.~\eqref{zak} is considered in the limit $k \to 0$. For the trivial phase with $\mu >+1$, the winding number is $\nu = -1/2$, whereas in the topological phase hosting MDMs  ($\mu <+1$), $\nu = 1/2$ (see Appendix A). The root of  the half-integral winding number is the following:  the divergence of $f_\alpha(k)$ as $k \to 0$ means that the curve trace by $\hat{n}_k$ covers a semicircle, rather than a full circle, as $k$ varies from $0 + \ep$ to $2\pi - \ep$). Although the topological invariant is a half-integer in both cases, the difference between the invariants in the two topologically different phases is unity, indicating that a topological phase transition separates the two half-integer quantized  phases \ct{viyuela16}.

\section{Long-range Kitaev Chain: Dynamics}

\label{sec_lrk_dynamics}

As we have discussed in the Sec.~\ref{cor_ee} that when the interaction is sufficiently long-ranged, the behaviour of the correlations functions in a gapped phases changes drastically i.e,. it has an purely algebraic or hybrid exponential-algebraic decay rather than an exponential decay. Also, a breaking of conformal symmetry occurs along a gap-less line accompanied by a violation of area law of EE in a gapped phase.  It is therefore would be quite interesting to investigate  the effect of the long-range interactions on the information propagation and equilibration of the local observables in the non-equilibrium scenario, e.g.,  after a sudden quench of
the chemical potential of the system.

\subsection{ Propagation of mutual information, growth of EE and equilibration after a sudden quench}

In a short-range interacting system, following  a sudden quench occurring  at $t=0$, there exists a {\it maximum} group velocity $v$ of the the quasiparticles such that there will be no correlations between two points separated by a distance $\Delta x$ up to a time $t=\Delta x/2v$ . This is given by the Lieb-Robinson bound~\cite{lieb72} which states there is an effective light-cone outside of which correlations get exponentially suppressed. The EE of a subsystem (say $A$) grows linearly in time after a quench~\cite{calabrese05,fagotti08} up to a time $\tau=\ell/2v$ (where $\ell$ is the  size of $A$), then it abruptly saturates to a constant value which proportional to $\ell$ and system thermalize to a generalized Gibbs ensemble (GGE)~~\cite{rigol07}. This behaviour of the EE after a sudden quench was explained  by Calabrese and Cardy in terms of a pair of entangled quasiparticles moving with opposite momenta~~\cite{calabrese05}. This can be understood as follows: following the sudden quench, a pair of entangled quasiparticles are produced at different points of the chain and propagate in opposite directions. The entanglement entropy receives a contribution from those pairs for which one quasiparticle lies inside the block while the other one lies out-side. Hence, the EE increases linearly in time till it saturates when the pair of quasiparticles that started in the middle of the block reaches its boundary.

\begin{figure}[]
	\centering
	\subfigure[]{
		\includegraphics[width=0.5\columnwidth,height=0.36\columnwidth]{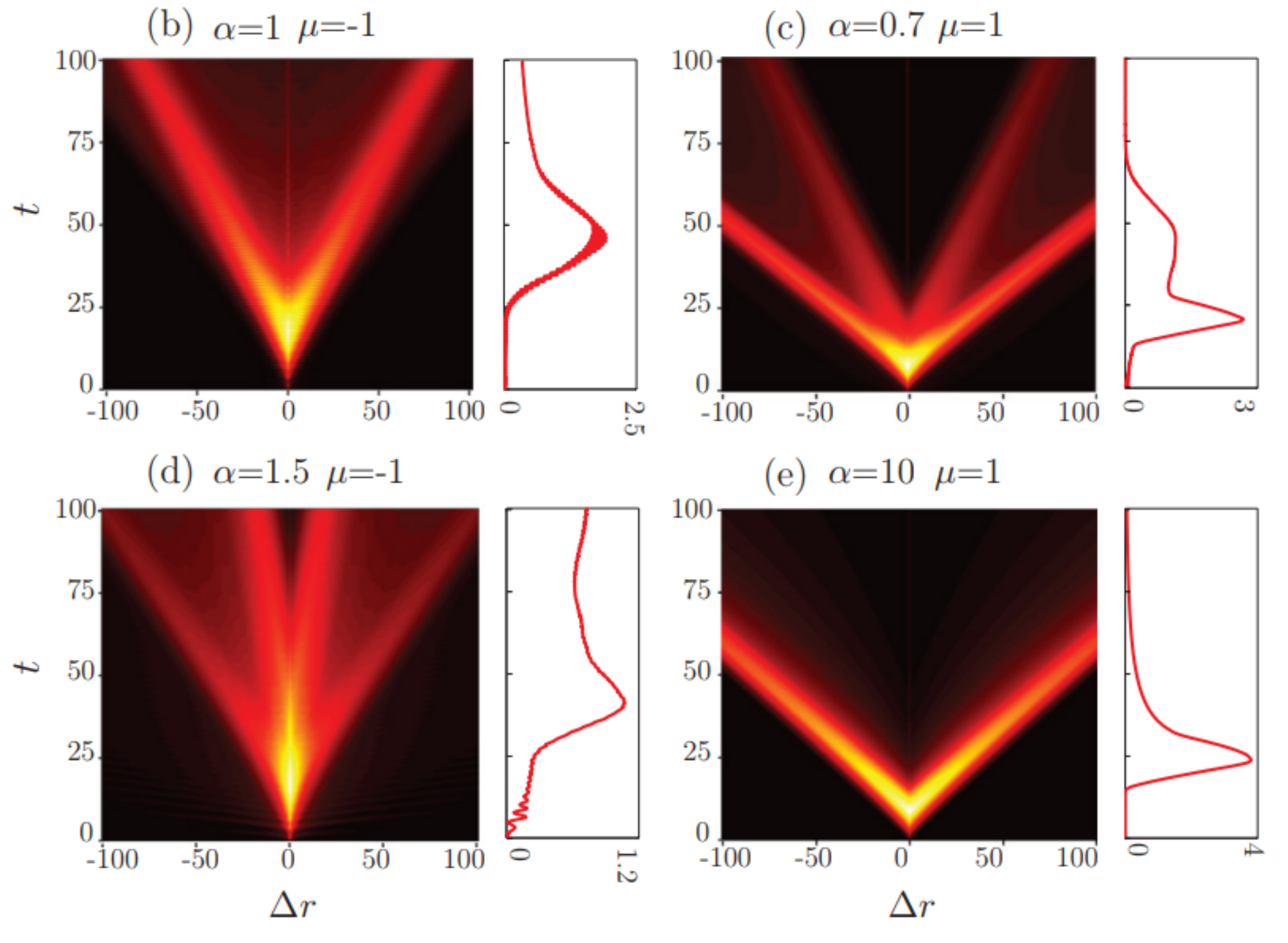}
		\label{mutual_info}}
	\hspace{0.05cm}
	\subfigure[]{
		\includegraphics[width=0.4\columnwidth,height=0.35\columnwidth]{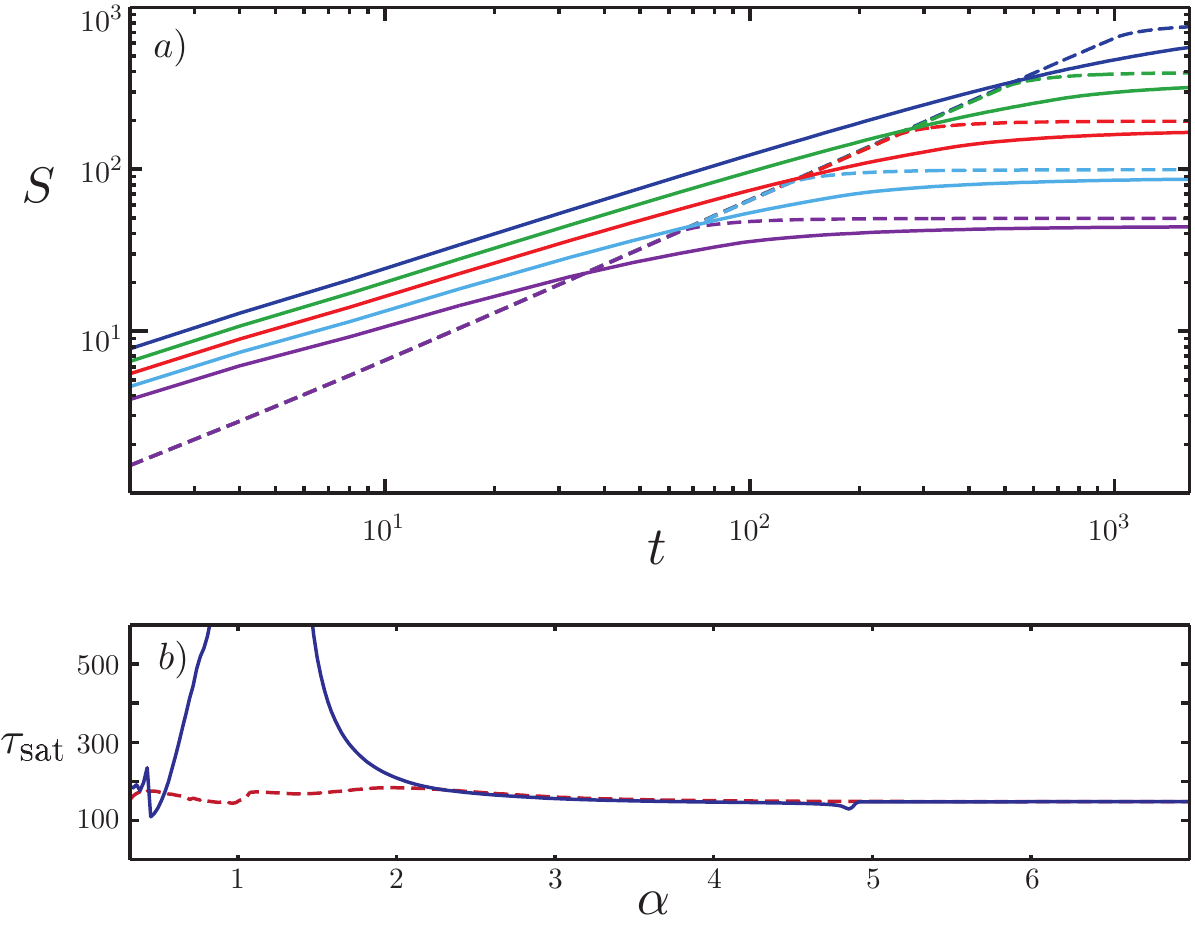}
		\label{ee_lrk}}
	\caption{(a) Propagation of mutual information between two subsystems separated by a distance $\Delta r$ in LRK chain after a sudden quench of the chemical potential from non-interacting ground state to critical points ($\mu=\pm1$) for several values of $\alpha$. The side panels show the mutual information propagation for a fixed value of $\Delta r$. (b) Time evolution of EE after a sudden quench from non-interacting ground state to $\mu=1$ for $\alpha=0.7$ with different values of system size. It shows an algebraic growth and very slow equilibration of the EE. The dashed lines represent  the same for the SRK chain ($\alpha=10)$ which shows   the usual ballistic growth of EE and an abrupt saturation at $\tau_{sat}=\ell/2$. The figure in the lower panel shows the saturation time of the EE $\tau_{sat}$ (defined as time takes to reach $95\%$ of the GGE entropy) as a function of $\alpha$, for $\mu=-1$ (blue solid line) and $\mu=1$ (red dashed line). (The figure taken from the Ref.~\cite{regemortel16} with permission from the authors and the publisher.)}
	\label{mutual_ee}
\end{figure}

In the case of a long-range interacting system correlations do not necessarily  obey the Lieb-Robinson bound and it may have immediate correlation between distance points~\cite{regemortel16,buyskikh16}.
In the Ref.~\cite{regemortel16}, Regemortel {\textit et al.} studied the propagation of mutual information between two parts (say $P$ and $Q$) of the LRK chain after a sudden quench from a product state. The mutual information is defined as,
\be
 \mathcal{I}_{PQ}=S_P+S_Q-S_{P,Q}, 
\ee
where $S_P$, $S_Q$ and $S_{P,Q}$ are the Von-Neumman entropy of the reduced density matrices for the region $P$, $Q$ and the composite region $P$ and $Q$, respectively. It has been reported that a small part off the mutual information build up immediately after the quench regardless the distance between two parts, but the largest build up of mutual information occurs within a well defined light-cone even for the sufficiently long-range interactions ($\alpha<1$). The most intriguingly, the long-range interactions slow down the local equilibration of the system as there is enough mutual information inside the time-like separations (see Fig.~\ref{mutual_info}). All these results has been explained in great details in terms of the quasiparticle group velocity distribution \cite{regemortel16}.  The slowing down of the local equilibration is also reflected on the time evolution of entanglement entropy after the sudden quench of the chemical potential. In contrast to the short-range interacting system, there is a power law growth of the EE; $S_\ell\sim t^\beta$ with $\beta<1$, in the case of LRK chain for $\alpha<1$ according to the Ref.~\cite{regemortel16} which is  in contradiction with the logarithmic growth  reported in ~\cite{vodola14}.

Let us note some related studies of the non-equilibrium dynamics of long-range transverse Ising chains.  For a long-range interacting  spin chain, the initial growth of the EE following a global sudden quench has been found to be logarithmic in time~\cite{schachenmayer13}.
For a local quench in the long-range Ising model in a transverse field, on the other hand, there exist three dynamical regimes: short-range-like with an emerging light cone for $ \alpha > 2$; weakly long-range for $1 < \alpha < 2$ without a clear light cone but with a finite propagation speed of quasiparticles; and fully non-local for $\alpha < 1$ with instantaneous transmission of correlations. This last regime breaks generalized Lieb-Robinson bounds. For any $\alpha$, the initial growth of the EE faster than logarithmic followed by a slow evolution in agreement with the appearance of a diffusive evolution  \cite{hauke13}. Thus, the behaviour of the growth of EE for a local quench in the long-range Ising model in a transverse field is similar to the case of LRK chain when $\alpha<1$ as discussed above. 

Finally, we conclude this section by a brief discussion on the \textit{prethermalization} observed in closed quantum systems with long-range ferromagnetic  Ising interactions (falling
as $1/r_{ij}^{\alpha}$, where  $r_{ij}$ is the distance  between between two spins)  both theoretically~\cite{worm13,kastner15,mori19} as well as experimentally~\cite{neyenhuis17}. The non-equilibrium dynamics of
  the spin system, which is non-integrable due to the presence of both a transverse ($h_z$) and longitudinal ($h_x$) fields, has been studied following a sudden quench \ct{mori19}.  In the presence of longitudinal field ($h_x\neq0$), for any $\alpha\in (0,1)$, the permutation operators {$P_{ij}$} of spins $i$ and $j$, are quasi-conserved quantity. The relaxation time of the expectation value of these permutation operators is not shorter than $O( L^{1-\alpha}) $ i.e., the relaxation timescale $\tau_{\text{rel}}\propto L^{1-\alpha}$. Another initial relaxation timescale ($\tau_{\text{ini}}$), associated with the growth of quantum correlations, has been evaluated by studying the time evolution of the spin-spin correlation functions as the initial stage of relaxation. It turns out, for any $\alpha$ in the range $0<\alpha<1$, the initial relaxation timescale goes as $\tau_{\text{ini}}\propto \ln L$ when the system size $L$ is sufficiently large. This logarithmic dependence of $\tau_{\text{ini}}$ on $L$ is because of the chaoticity of the underlying classical dynamics where the quantum fluctuations grow exponentially fast as $e^{\kappa t}/\sqrt{L}$ with positive constant $\kappa$ .  Therefore, there exists a large timescale separation between the two steps of relaxation and thus prethermalization occurs in the system for any $\alpha\in (0,1)$. In absence of the longitudinal field ($h_x=0$), prethermalization occurs only for $0<\alpha<1/2$~\cite{worm13,kastner15}. This is because of
  the fact that the underlying classical dynamics is not chaotic rather regular with $\tau_{\text{ini}} \propto L^{\gamma}$ with $\gamma=\text{min}\left\{1/2,1-\alpha\right\}$ instead of the logarithmic in $L$.

For a periodically driven many body spin system with long-range disordered interactions, the disorder averaged energy absorption rate at high temperatures has been found to decay exponentially with the driving frequency and there exist a prethermal plateau in which dynamics is governed by an effective, static Hamiltonian for long times~\cite{ho18}.

\subsection{Kibble-Zurek scaling}

 The LRK chain has also been studied in the context defect generation following a ramp of the chemical potential term $\mu$ is tuned as $\mu(t)={{t}/{\tau_Q}}$, from a large negative value to a final value $\mu=0$
 \ct{anirban17}.
 When a quantum many-body system is quenched across a QCP by changing a parameter (say, linearly as $t/\tau_Q$), the dynamics is necessarily non-adiabatic due to
the diverging relaxation time associated with the QCP; this results in the production of defect (or excess energy) in the {\it final} state of the system reached following the quench \cite{Kibble76,Zurek96}.
Remarkably, the density of defect ($n_d$) exhibits a universal scaling with the inverse quenching rate $\tau_Q$;  this is known as the Kibble-Zurek (KZ) scaling which has been studied extensively in recent years \cite{ZDZ05,Polkovnikov05,dziarmaga05,Damski05,damski_zurek06,mukherjee07,divakaran08,sen08,shreyoshi08}.  (For review, see [\onlinecite{dziarmaga10,polkovnikov11,dutta15}]).  Following a slow passage across  
an isolated critical point, the density of defect  in the final state satisfies the scaling relation \ct{dziarmaga05}:

\be
n_d \sim \frac 1 {{\tau_Q}^{{\tilde \nu} d/{\tilde \nu} z +1}},
\label{eq_kz}
\ee
where ${\tilde \nu}$ is correlation length exponent and $z$ is the dynamical exponent associated with the QCP across which the system is quenched. In the subsequent discussion, we shall consider
the LRK for the parameter $\alpha >1$.

Let us initially prepare the LRK chain in ground state with a large negative $\mu$ to which and the ramp is subjected to.  {Referring to the quasi-momentum mode $k$,  the state of the system at any instant $t$ can be written as  $|\psi_k(t)\rangle=u_k(t)|0\rangle+v_k(t) 
|1\rangle$, where $|0\rangle$ and $|1\rangle$ are the diabatic basis states $(1,0)^T$ and $(0,1)^T$ with reference to ($2\times2$) Hamiltonian in Eq.~\eqref{ham_k}. The corresponding Schr\"{o}dinger equation
can be written in the form:}
{
\begin{eqnarray}
i\dfrac{d }{dt}u_k(t)=-(w\cos k+\mu(t))u_k(t)+i \Delta f_{\alpha}^{\infty}(k)v_k(t)\nonumber\\
i\dfrac{d }{dt}v_k(t)=-i\Delta f_{\alpha}^{\infty}(k)u_k(t)+(w \cos k+\mu(t))v_k(t),
\label{scheq}
\end{eqnarray}
with the initial condition $|u_k|^2=1$ and $|v_k|^2=0$. Using the transformation,  $\tau=i\tau_Q\Delta f_{\alpha}^{\infty}(k)\left(t/\tau_Q-w\cos k\right)$, we can recast  the above Eq.~(\ref{scheq}) to the standard  Landau-Zener (LZ)  form \ct{landau,sei,vitanov}
\ {
\begin{eqnarray}
i\dfrac{d }{d\tau}u_k(t)&=&-(\tau\tilde{\Delta}_k)u_k(t)+v_k(t)\nonumber\\
i\dfrac{d }{d\tau}v_k(t)&=&u_k(t)+(\tau\tilde{\Delta}_k)v_k(t)
\label{lzeq}
\end{eqnarray}
where $\tilde{\Delta}_k=\left(\tau_Q(\Delta f_{\alpha}^{\infty}(k))^2\right)^{-1}$.}}
{For slow passage through the QCP,  the excitation probability  can be calculated using the LZ non-adiabatic transition probability \ct{landau,sei} that
the system ends in the excited state of the final Hamiltonian $\mu=0$;
\begin{eqnarray}
p_k= e^{-\pi/\tilde{\Delta}_k}\simeq e^{-\pi\tau_Q(\Delta f_{\alpha}^{\infty}(k))^2}.
\label{lztp}
\end{eqnarray}}
\noindent In the large $\tau_Q$ limit, $p_k$ will be significant only for the modes close to the critical mode $k_c=0$ and  hence we use the expansion formula of the poly-logarithmic functions in the limit $k\to 0$,
considering three limiting situations and calculate the corresponding scaling of the defect density:

\textbf{Situation I:} $\alpha\not\in\mathbb{Z}$; in this case, one can use the expansion:

\begin{eqnarray}
(f_{\alpha}^{\infty}(k))^2&=&4\cos^2{{\pi\alpha}\over{2}}\Gamma^2(1-\alpha)k^{2(\alpha-1)}+4\zeta^2(\alpha-1)k^2\nonumber\\&&+8\cos{{\pi\alpha}\over{2}}\zeta(\alpha)\Gamma(1-\alpha)k^{\alpha}+o(k^3)\non\\
&=&c_1(\alpha)k^{2(\alpha-1)}+c_2(\alpha)k^{\alpha}+c_3(\alpha)k^{2}+o(k^3)\non\\
\label{polyee1}
\end{eqnarray}
 {where we have used standard Gamma functions($\Gamma$) and Riemann zeta functions($\zeta$) \cite {vodola16,functions}  \  {and the coefficients are $c_i(\alpha)$s given by:}
	\begin{eqnarray}
	c_1(\alpha)&=&4\cos^2{{\pi\alpha}\over{2}}\Gamma^2(1-\alpha),\\
	c_2(\alpha)&=&8\cos{{\pi\alpha}\over{2}}\zeta(\alpha)\Gamma(1-\alpha),\\
	c_3(\alpha)&=&4\zeta^2(\alpha-1).
	\end{eqnarray}}
Inspecting Eq.~\eqref{polyee1}, we find that for $\alpha <2$, the term $k^{2(\alpha-1)}$ dominates as $k \to 0$. Hence, the density of quasiparticle excitation in the final state  at the end of the drive  in the thermodynamic limit   can be calculated as
\begin{eqnarray}
n_d={{1}\over{2\pi}}\int_{-\pi}^{\pi}p_k dk \approx {{1}\over{2\pi}}\int_{-\infty}^{\infty} e^{-\pi\Delta c_1(\alpha)\tau_Q k^{2(\alpha-1)}} dk 
={{1}\over{2\pi}}\left[{{\Gamma({{1}\over{2\alpha-2}})}\over{(\pi\Delta c_1(\alpha)\tau_Q)^{{1}\over{2\alpha-2}}}}\right],
\label{kzf1}
\end{eqnarray}
where we have extended the range of integration over $k$ from $-\infty$ to $\infty$. 	Analysing the spectrum in Eq.~\eqref{clrspectrum}, we find that in this case, the dynamical exponent $z=(\alpha-1)$, while
${\tilde \nu} z=1$; we therefore find that the scaling of $n_d$ in Eq.~\eqref{kzf1} obtained through the Landau-Zener transition  formula is consistent with the Kibble Zurek prediction in Eq.~\eqref{eq_kz}.
On the other hand, for $\alpha >2$, the term $k^2$ dominates and hence we retrieve the scaling for the short-range situation $n_d \sim 1/{\sqrt \tau}$ with ${\tilde \nu}=z=1$ \ct{dziarmaga05}.

\begin{figure}[]
	\centering
	\subfigure[]{
		\includegraphics[width=0.35\columnwidth,height=0.3\columnwidth]{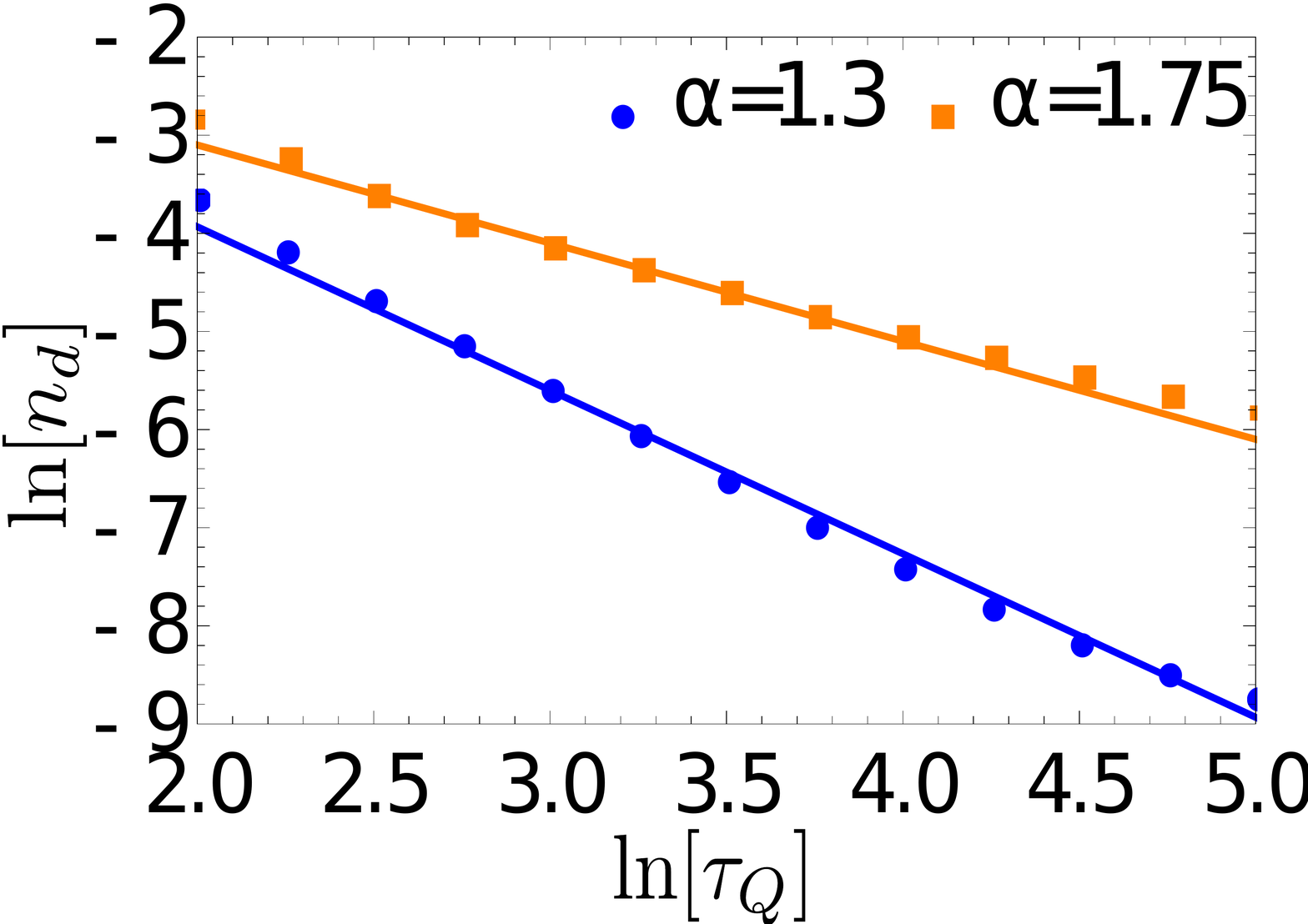}
		\label{kz1}}
	\hspace{0.05cm}
	\subfigure[]{
		\includegraphics[width=0.35\columnwidth,height=0.29\columnwidth]{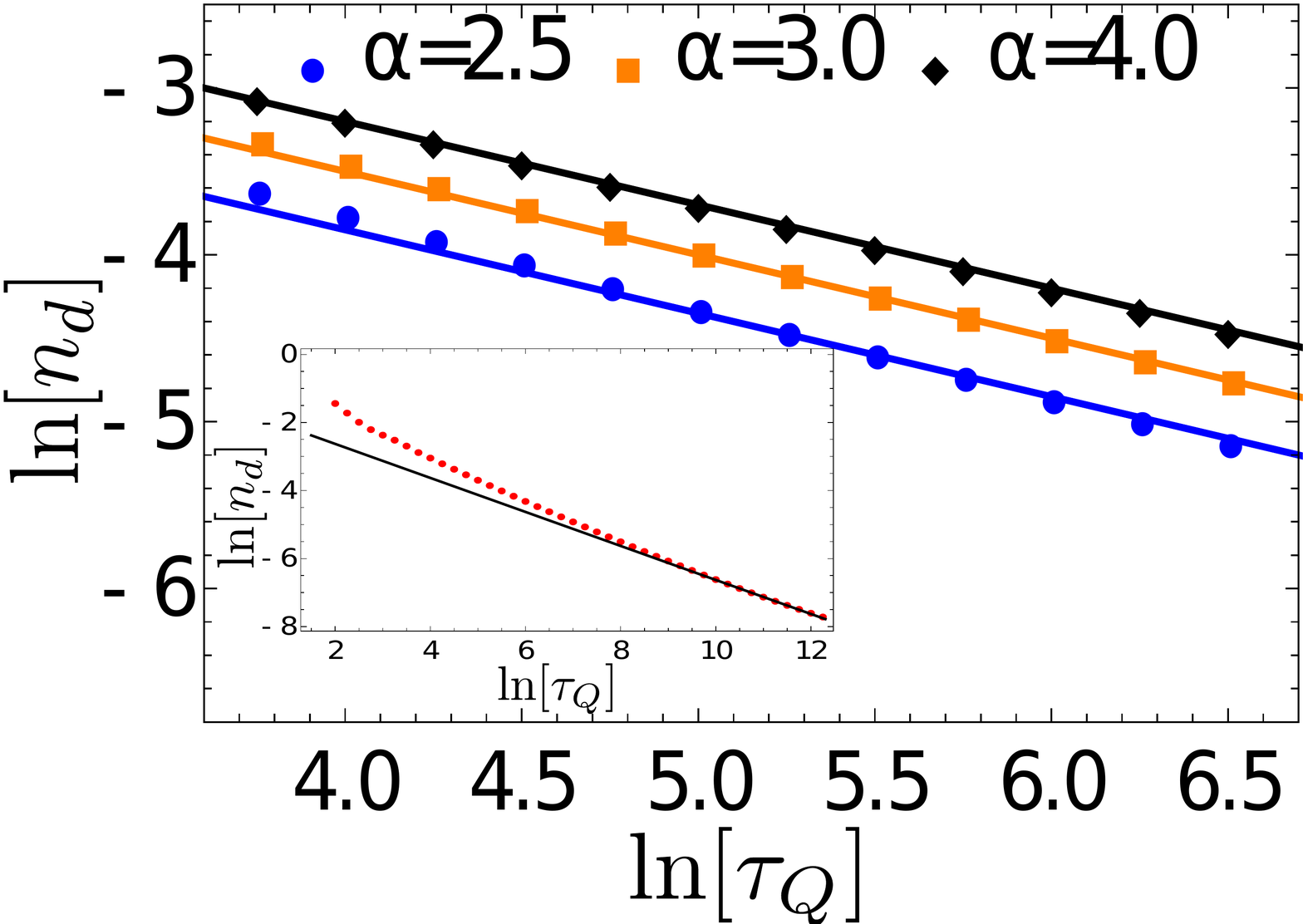}
		\label{kz3}}
	\caption{The variation of the defect density $n_d$ with inverse quenching rate $\tau_Q$:
		{\bf (a)}  The figure shows that the  scaling exponent (determined by the slope of the curve) depends on the  interaction range $\alpha$ for $\alpha <2$. The solid lines show the  theoretically predicted values $1/(2\alpha -2)$  and the dotted lines are from numerical {solution of the differential Eqs. \eqref{lzeq} using Runge-Kutta method with $L=10000$, $\mu_i=-10$ and $\mu_f=0$.}
		{\bf (b)} The scaling exponent does no longer  change with the range of interaction  for $\alpha \ge 2$, rather gets saturated to  the short-range value $1/2$. In the marginal case $\alpha=2$, the exponent reaches the short-range limit of $1/2$ only in the asymptotic limit as shown in the inset. (Figure taken from [\onlinecite{anirban17}].)}
	\label{kz}
\end{figure}

For integer $\alpha$, the expansion formula for poly-logarithmic functions  is given by 
\begin{eqnarray}
f_{\alpha}^{\infty}(k)&=&-\frac{i^\alpha k^{\alpha-1}}{(\alpha-1)!}\left[(1+(-1)^\alpha)(\mathbb{H}_{\alpha-1}-\ln{k}-\frac{i\pi}{2})+i\pi\right]\nonumber\\&&-2\sum_{\substack{m\in odd \\ m\neq \alpha-1}}\frac{\zeta(\alpha-m)i^{m+1}}{m!}k^m,
\label{polyee2}
\end{eqnarray}
where $\mathbb{H}_n$ is nth Harmonic number. Using this expansion, we shall probe the following cases:\\

\textbf{Situation II: } $\alpha$ is an integer; for any  $\alpha\neq2,\in\mathbb{Z}$ one finds: 
\begin{eqnarray}
(f_{\alpha}^{\infty}(k))^2&=&4(\zeta(\alpha-1))^2 k^2+o(k^3).
\label{pkintk}
\end{eqnarray}
{Using Eq.~\eqref{pkintk}, one can similarly  calculate the  scaling of the density of quasiparticle excitation in the final state
\begin{eqnarray}
n_d&\simeq&{{1}\over{(4\pi(\zeta(\alpha-1))^2 \tau_Q)^{{1}\over{2}}}}. 
\label{kz2}
\end{eqnarray}
Thus when the interaction is sufficiently short-ranged (i.e., $\alpha >2$), we again retrieve the short-range KZ scaling.

\textbf{Situation III:} Marginal case $\alpha =2$; the expansion of $(f_{\alpha}^{\infty}(k))^2$ for $\alpha$=2 can be calculated from the expansion Eq.(\ref{polyee2}):
\begin{eqnarray}
(f_{2}^{\infty}(k))^2&=&4(1-\ln k)^2 k^2+o(k^3);
\label{pk2k}
\end{eqnarray}

\noindent We note that for $\alpha$=2,  there is a prominent  logarithmic correction in the leading order to the expression of $p_k$ which leads to sub-leading corrections to the scaling of $n_d \sim 1/\sqrt{\tau_Q}$. However,
in the asymptotic limit  of $\tau_Q \to \infty$, the sub-leading corrections drops off  yielding the short-range scaling relation.

We thus have the KZ scaling exponent $1/(2\alpha -2)$,  which reduces to the short-range value when $\alpha \to 2$ while for  $\alpha >2$  the   scaling exponent  gets saturated to $1/2$.  This establishes that
the case $\alpha=2$ marks the boundary between the long-range and the short-range behaviour so far as the KZ scaling is concerned.  Furthermore, in the marginal case $\alpha=2$, there are
non-universal sub-leading corrections which vanish in the limit of $\tau_Q \to \infty$.  It is remarkable that the situation with $\alpha=2$, which is the marginal situation for the classical Ising
chain, turns out to be marginal also in the KZ scaling. These KZ scaling relations  predicted analytically have also been numerically verified as shown in Fig.~\ref{kz}. We conclude this section
with the note that KZS has been also explored for a long-range transverse Ising chain using numerical techniques and the corresponding scaling exponent is found depend non-trivially
on the range of the interaction \ct{jaschke17}: For example, in  the ferromagnetic case a  slower increasing defect density has been observed  in comparison to the nearest-neighbor limit when increasing the quench velocity.

\section{Some related works on the dynamics of the LRK chain}
\label{sec_related}

\noi \textbf{(i) Periodically kicked LRK chain:}  The non-equilibrium dynamics of the LRK chain subjected to a periodic $\delta$-kicks in the chemical potential, 
\be
\mu(t)=\mu+V\sum_{n=-\infty}^{\infty} \delta(t-nT),
\ee
where $V$ and $T$ is the kicking strength and time period, respectively, has been recently studied~\cite{bhattacharya19}. The quasienergy spectrum of the Floquet Hamiltonian have been studied as function of $\mu$ for various values of the parameters $\alpha$ and kicking frequency ($\omega=2\pi/T$). It has been shown that the periodic $\delta$-kicks generates new massless Majorana modes (for $\alpha>1$) and massive Dirac modes (for $\alpha<1$) at the quasienergies $0$ and $\pi$ in addition to the modes  occurring at  zero energy in the topological phase of the undriven chain. Moreover, these end modes are separated from the bulk by finite gaps and are therefore topologically protected. The critical values of the chemical potential $\mu$ can be calculated as function of $\omega$ and $V$, where the massless Majorana and massive Dirac modes occur. Interesting, both massless and massive modes can co-exist for certain values of $\mu$ if we decrease the kicking frequency. The existence of the these modes has been characterised by an appropriate bulk topological invariant in the periodic boundary condition.    \\

\noi \textbf{(ii) Periodically driven LRK chain:} Periodically driven LRK chain has also been studied in Ref.~\cite{nandy18} in the context of equilibration and  mutual information propagation within the framework of
Floquet theory;  in this work, the chemical potential   $\mu$  is subjected
to periodic pulses and sinusoidal variations.  It has been observed that all the local observables confined in a subsystem (with size $\ell<<L$) synchronizes with the driving frequency $\omega$ and reaches a periodic steady state in the asymptotic limit  (i.e.,  the number of stroboscopic periods $n\rightarrow\infty$) irrespective of the value of the exponent $\alpha$ and the frequency $\omega$.  To study the equilibration process,
an appropriate distance measure $\mathcal{D}_{\ell}(n)$ ($\in \left[0,1\right]$) (between the instantaneous correlation matrix and the asymptotic correlation matrix) has been studied for different values of $\alpha$ and $\omega$. Interestingly, there is a critical value of the exponent $\alpha_c$, above which two different dynamical phases exist depending on the frequency of driving. More specifically, for $\alpha>\alpha_c$, the distance measure decays as $\mathcal{D}_{\ell}(n)\sim (\omega/n)^{3/2}$ when $\omega\rightarrow\infty$ and $\mathcal{D}_{\ell}(n)\sim (\omega/n)^{1/2}$ when $\omega\rightarrow 0$, which indicates at least one dynamical phase transition occurs at some frequency $\omega_c$ similar to the short-range system~\cite{sen16}. On the other hand no such dynamical phase transition present for $\alpha<\alpha_c$ as $\mathcal{D}_{\ell}(n)\sim (\omega/n)^{1/2}$ for any value of $\omega$. Under the application of periodic driving, propagation of mutual information between two subsystems has been also studied as function of $\alpha$ and $\omega$. The signature of the dynamical phase transition is also observed in the mutual information propagation. It is reported that for $\alpha>2>\alpha_c$, mutual information propagates in a well defined single light-cone (like the short-range intereacting system) when $\omega>\omega_c$; on the other hand, it shows multiple light-cone like features when $\omega<\omega_c$. Furthermore, when $\alpha\leq2$, system does not exhibits any light-cone like structure and information can propagate instantaneously throughout the system similar to the case after a quench~\cite{regemortel16,lepori17, buyskikh16}. \\

\noi \textbf{(iii) Dynamical Quantum Phase transitions (DQPTs):} DQPTs, introduced by Heyl {\it et~al.} \ct{heyl13} (see also \ct{pollmann10,heyl14,vajna14,heyl15}, are non-equilibrium quantum phase transitions occurring in the subsequent temporal evolution of a quenched system at different instants of real time and are manifested in non-analyticities in the Loschmidt echo ($\mathcal {L}$).  Here, a quantum many-body system is initially prepared in the ground state $|\psi_0 \rangle$ of the initial
Hamiltonian $H(\lambda_i)$ and the parameter $\lambda_i$ is suddenly quenched to a final value $\lambda_f$.  DQPTs occur at those instants of time (denoted by critical times $t_c$) at which  the evolved state $|\psi(t) \rangle = \exp(-i H(\lambda_f)t) |\psi_0 \rangle$ becomes orthogonal to the initial state so that ${\mathcal L}(t)= |\langle \psi_0|\exp(-i H(\lambda_f)t) |\psi_0 \rangle|^2 $ vanishes. For an integrable model,
reducible to decoupled  $2 \times 2$ problems for each momenta mode, $t_c^n =  ({\pi} /{2\epsilon_{k_*}^f})  (2n+ 1 )$, where $k_*$ is the critical mode for which the overlap vanishes,
 $2\epsilon_{k_*}^f$ is the corresponding energy gap of the final Hamiltonian and $n=0, \pm1, ...$~.  For a one-dimensional system, the corresponding dynamical free energy density ${\mathcal F}= - (1/L)\ln \langle \psi_0|\exp(-i H(\lambda_f)t) |\psi_0 \rangle$ (or the rate
function of return probability), where $L$ is the chain length, shows cusp singularities  at those values of $t_c$. These non-analyticities  have also been observed following a preparation of the initial state through a linear ramping of the parameter
$\lambda$ \ct{sharma16} also in non-integrable models \ct{karrasch13,kriel14,sharma15}. Further, these non-analyticities have been experimentally detected \ct{jurcevic16,flaschner16}. Although there is no local order 
parameter, these non-equilibrium transitions can be characterised by  a dynamical topological order parameter which is constructed from the Pancharatnam phase extracted from the Loschmidt overlap
amplitude  $\langle \psi_0|\exp(-i H(\lambda_f)t) |\psi_0 \rangle$ \ct{budich15} and shows an integer jump of unit magnitude at every critical time. In two dimensional situations, the non-analyticities manifest in the time derivative of the dynamical free energy density
\ct{schmitt15,vajna15,utso16} and the associated  topological winding number has been derived \ct{utso17}. DQPTs have also been explored  when the initial state is a mixed state \ct{bhattacharya17,heyl17} and also for open quantum systems \ct{bandyopadhyay17}.

Similar DQPTs have been probed for the LRK chain following a sudden change in the chemical potential $\mu$ \ct{anirban17}. Interestingly,  the existence of a new region in the $\mu-\alpha$ plane  has been established; in this region, one  finds three instants of cusp singularities in the dynamical free energy  associated with a single sector of  zeros (i.e., for a single $n$) of the Loschmidt overlap amplitude. Notably, the width of this region shrinks as $\alpha$ increases and vanishes in the limit $\alpha \to 2$ indicating that this special region is an artefact of long-range nature of the  LRK Hamiltonian \ct{anirban17}. This study also shows that $\alpha=2$
marks the boundary between the short-range and the long-range limit so far as dynamics is concerned.

It is note-worthy that the long-range interacting quantum Ising chain has been studied in connection to two dynamical phase transitions \ct{zuncovic16, halimeh17,halimeh171,homrighausen17}; Ð one  based on an
order parameter in a (quasi-)steady state, and the other based on non-analyticities (cusps) discussed above. The long-range Ising chain has also been studied 
using a cluster mean field theory and the range of interaction has been found to play a dominant role, at least for the small values of $\alpha$ \ct{piccitto19}.\\

\noi \textbf{(iv) Survival probability of edge Majorana modes:} The  temporal evolution of a zero energy edge Majorana present in the topological phase of a  SRK chain has also been studied following a sudden quench of the chemical potential $\mu$ \ct{rajak14}.  The system
is quenched either to the topologically trivial phase or to the QCP separating these two phase and the survival probability of one of the edge Majoranas is observed as a function of time. Interestingly when the chain is quenched to the QCP, one find a nearly perfect oscillations of the survival probability, indicating that the Majorana travels back and forth between two ends, with a time period that scales with the system size. This
is manifested in the Majorana survival probability (or the Loschmidt echo) which is the modulus square of the overlap between the initial Majorana state and its time evolved counterpart evolving with the final Hamiltonian; these oscillations, when quenched to the QCP, 
can be attributed to the  to the nearly equi-spaced nature of the bulk modes and linearity of the spectrum at the QCP. We note that similar  quasi-periodic oscillations of edge states 
have also been observed in the case of topological insulators \ct{patel13} and Chern insulators \ct{sacramento14}.  Recently, a similar quenching  to the QCP  of  the LRK chain  has been studied. For a finite size system,
it has been found that the revivals in the Loschmidt echo are connected to the energy gap. 
Interestingly, when  long-range superconducting term dominates, 
the first revival time (periodicity) is found to scale inversely with the group velocity at the gap closing point,
instead of the maximum group velocity \ct{mishra18}.\\

\section{Conclusion}
\label{sec_conclusion}
In this review article, we have discussed long-range interacting classical and quantum Ising chains and their critical properties. We have pointed out that there exists a special type of BKT transition
in inverse square Ising chain and the corresponding field theoretical study is still lacking. At the same, interestingly any finite temperature transition of its quantum counterpart belongs to the
universality class of the classical  BKT transition and thereby yielding a line of BKT transitions in the $h$-$T$ plane. We have also mentioned some recent studies on the dynamics of long-range quantum
Ising chains in subsequent discussions. It is note-worthy that  long-range  interacting spin models have also been studied in the context of the non-equivalence of ensembles
and the diverging equilibration time \ct{kastner10,kastner11}.

The major emphasis of the review, however, lies on the recent developments of 
the LRK chain. While the long-range interacting Ising chains are non-integrable, the LRK chain is integrable and can be reduced to decoupled two level problems; further, even in the
presence of long-range pairing terms, the LRK chain respects all the symmetries of the BDI class and hence its topological properties can be characterised by a winding number. Regarding
the topological phase diagram of the model, the long-range interaction plays a dominant role when the decay constant $\alpha <1$;  both  the topological phase and the non-topological 
phase are characterised by half-integer winding numbers, which however, shows a jump of unit magnitude across the topological transition. However, for $\alpha >1$, the topological
behaviour of the LRK chain is similar to that of the SRK chain  (with $\Delta \geq 0$) with  a topological phase ($\nu=1$) and a topologically trivial phase  ($\nu=0$) separated
by QCP.  For the LRK chain with an open boundary condition and  $\alpha >1$,  symmetry protected massless edge MZMs exist as in   case of the SRK model.  On the other hand,
when the long-range pairing dominates (i.e., $\alpha <1$), these edge MZMs hybridise to form topologically protected massive Dirac modes. The occurrence of these massive  modes is a cardinal feature of the LRK chain for $\alpha <1$. We therefore conclude that the line $\alpha=1$ marks the crossover between the long-range and the short-range topological behaviour. 

The out of equilibrium of the LRK chain is also influenced by the long-range nature of interaction.  The ballistic nature of the growth of the  EE following a sudden quench  disappears for $\alpha <1$, rather
there is a power-law (or logarithmic) growth to asymptotic steady value that obeys the volume law.   In the context of the mutual information propagation, it has been observed that even though a light cone
like propagation is evident is short-range limit, in the other situation there is an instantaneous propagation of information and the light cone structure  disappears. For the sudden quenching case, the
the light cone disappears for $\alpha <3/2$ as the group velocity of the quasiparticles diverge at $k=0$.   In the case of periodically driven
LRK chain, on the other hand, the same happens for $\alpha <2$ because the stroboscopic group velocity diverges. 
Also in DQPTs those occur following a sudden quench and  in the KZS scaling following a linear ramp, it appears that the long-range interactions play a dominant role only  if $\alpha <2$.
Recently, the Lieb-Robinson bounds for  quantum systems coupled to a bath  with long-ranged interactions has also been studied \ct{sweke19}.

Let us also comment on some studies on the higher dimensional version of the models discussed in this review.  Using  matrix product states and density-matrix renormalization-group algorithms,  the ground-state phase diagram of the long-range interacting triangular Ising model with a transverse field on six-leg infinite-length cylinders has been studied and it has been suggested that the long-range quantum fluctuations always lead to long-range correlations, where correlators exhibit power-law decays instead of the conventional exponential drops observed for short-range correlated gapped phases \ct{saadatmand18}. A two-dimensional transverse Ising model in the presence of algebraically decaying long-range ferromagnetic/antiferromagnetic interactions residing on the square and triangular lattice has been studied using a numerical method  that combines the  perturbative continuous unitary transformations with classical Monte Carlo simulations.  It has been found that  the unfrustrated systems change from mean-field to nearest-neighbor universality with critical exponents continuously varying  with the range parameter.    In the frustrated case,  the system sticks to  the universality class of the nearest-neighbor model independent of the long-range nature of the interaction for the square lattice.  On the contrary, for the triangular lattice the quantum criticality is terminated by a first-order phase transition line \ct{fey19}.

We note that  it has been recently proposed that an experimental platform based on a planar Josephson junction comprising of a two dimensional degenerate electron gas with spin-orbit coupling, can host Majorana modes \ct{pientka17}.  This set up has been shown to be equivalent to a long-range Kitaev chain \ct{liu18}. Further when a magnetic impurity is placed on top of a conventional $s$-wave superconductor, a Shiba state with energy inside the superconducting gap is formed when the interaction between the impurity and the quasiparticles in the bulk superconductor is sufficiently strong \ct{shiba68}. A periodic one dimensional array of such impurities can be described by a LRK chain as long as the chain length is small compared to the coherence length of the host superconductor \ct{hoffman16}.

Motivated by the experimental studies of two-dimensional topological superconductors \ct{menard17}, the phase diagram and the edge states of a two-dimensional  LRK chain with both algebraically decaying hopping and the superconducting pairing terms has been explored \ct{viyuela18}.  Two topological phases with different chirality
and one  non-topological phase, characterised by a topological winding number, have been found. Topological phases are characterised by propagating edge Majorana
modes. When both  the hopping and the paring terms decay with the same 
exponent, one of the topological phases gets significantly enhanced. On the contrary, when the decay of the pairing term is slower than the hopping term, a new phase
emerges in which the edge Majoranas couple non-locally and remain gapped even in the thermodynamic limit; remarkably, these non-local edge states are still separated from
the bulk by a gap and are localized at both the edges at the same time.

\begin{acknowledgments}
We acknowledge  Jaynata K. Bhattacharjee, Anirban Dutta and Diptiman Sen for collaboration in related works. We  also acknowledge Souvik Bandyopadhyay and  Sourav Bhattacharjee  for  discussions. We are
grateful to Davide Vodola and Mathias Van Regemortel for allowing us to reuse figures from their papers. We are also thankful to American Physical Society for granting permission
for the same figures.
\end{acknowledgments}

\appendix

\section{Calculation of winding number for LRK chain}\label{app_wn}

In this Appendix, we shall explicitly  calculate the winding number ($\nu$) as  defined in Eq.~\eqref{wn} for the LRK chain using  $H_k$ as in  Eq. \eqref{Hlrk} for different values of $\alpha$. Let us recall 
\be
\nu~=~\frac{1}{2\pi}\oint d\phi_k~=~\frac{1}{2\pi}\int_{0}^{2\pi}dk \left(\frac{d\phi_k}{dk}\right),
\label{wna}
\ee
where the angle $\phi_k=\tan^{-1}\left[f_{\alpha}(k)/(\mu+\cos k)\right]$ for the Hamiltoninan $H_k$ and the closed line integral is over the Brillouin zone.

Let us first consider the case $\alpha\rightarrow\infty$ (i.e., the SRK chain with $\Delta=\omega=1/2$):  the band touching is only possible at momenta values $k=0$ and $k=\pi$ in the Brillouin zone which only contribute in the integral in Eq.~\ref{wna}.  and we will linearise  $\phi_k=\tan^{-1}\left[\sin k/(\mu+\cos k)\right]$ around these two points. Linearising around the point $k=0$, we get $\phi_k=\tan^{-1}\left[k/(\mu+1)\right]$ and thus
\be
\frac{d\phi_k}{dk}=\frac{(1+\mu)}{(1+\mu)^2+k^2}.
\ee
The contribution to the $\nu$ from $k=0$ is then
\be
\nu_1=\frac{a}{2\pi}\int_{-\infty}^{\infty}\frac{dk}{a^2+k^2},
\ee
where $a=1+\mu\neq0$.  Using the continuum approximation,  we change the limit of integration  from $-\infty$ to $\infty$ which yields
\ba
\nu_1= \begin{cases}
	+\frac{1}{2}, & \text{for } \mu > -1, \\
	-\frac{1}{2}, & \text{for } \mu<-1.
\end{cases}
\ea
Similarly, linearising around  $k=\pi$ yields  $\phi_k=\tan^{-1}\left[(\pi-k)/(\mu-1)\right]$ and thus,
\ba
\nu_2= \begin{cases}
	+\frac{1}{2}, & \text{for } \mu < 1, \\
	-\frac{1}{2}, & \text{for } \mu > 1.
\end{cases}
\ea
Therefore, the winding number  for the SRK chain is given by,
\ba
\nu=\nu_1+\nu_2=\begin{cases}
	1, & \text{for } \lvert\mu\rvert < 1, \\
	0, & \text{for } \lvert\mu\rvert > 1.
\end{cases}
\ea
Let us now proceed to the long-range situation: for any finite value of $\alpha$ ($\alpha>1$), band gap closes at both the point $k=0$ and $k=\pi$ and leading order expansions in $k$ (or ($\pi-k$)) are the same as the SRK model with different positive coefficient. Therefore, the winding number happens to be the same as SRK model for any $\alpha>1$.  

On the other hand, when $\alpha<1$, the band touching is only possible at $k=\pi$ (as $f_{\alpha}(k)\rightarrow\infty$ as $k\rightarrow 0$ for $\alpha<1$) and $\nu$ has contribution only from the point $k=\pi$. Linearising the function $f_\alpha(k)$ about the point $k=\pi$,
\be
f_{\alpha}(k)\rvert_{k=\pi}\sim \left[-\frac{\partial}{\partial k}\left( \text{Li}_\alpha(e^{ik})\right)\right]_{k=\pi}\left(\pi-k\right),
\ee
and hence,  $\phi_k=\tan^{-1}\left[g(\alpha)(\pi-k)/(\mu-1)\right]$ where $g(\alpha)=\left[-\frac{\partial}{\partial k}\left( \text{Li}_\alpha(e^{ik})\right)\right]_{k=\pi}$. Since $g(\alpha)$ is a real positive quantity for $0\leq\alpha<1$. Therefore, the winding number is given by,
\ba
\nu=\frac{1}{2\pi}\int_{-\infty}^{\infty}\frac{g(1-\mu)dk}{(\mu-1)^2+g^2\left(\pi-k\right)^2}=\begin{cases}
	+\frac{1}{2}, & \text{for } \mu < 1, \\
	-\frac{1}{2}, & \text{for } \mu > 1.
\end{cases}
\ea
Therefore, the long-range interactions manifest itself in the topological winding number $\nu$ characterising the LRK chain for $\alpha \le 1$ (see Fig.~\ref{lrpd} and related discussions in the main text)


\begin{thebibliography}{11}

\bi{rulle68} D. Ruelle, Commun. Math. Phys. {\bf 9}, 267 (1968).
\bi{dyson69} F. J. Dyson, Commun. Math. Phys. {\bf 12}, 91 (1969); Commun. Math. Phys. {\bf 12}, 212 (1969).
\bi{kac69} M. Kac and C. J. Thompsom, J. Math. Phys. {\bf 10}, 1373 (1969).

\bi{thouless69} D. J. Thouless, Phys. Rev. {\bf 187}, 732 (1969).

\bi{fisher72} M. E. Fisher, S. K. Ma and B. G. Nickel, Phys. Rev. Lett {\bf 29}, 917 (1972).

\bi{wu91} W. Wu, B. Ellman, T.F. Rosenbaum, G. Aeppli, and D.H. Reich
Phys. Rev. Lett. {\bf 67}, 2076 (1991); W. Wu, D. Bitko, T.F. Rosenbaum, and G. Aeppli, Phys. Rev.
Lett. {\bf 71}, 1919 (1993).

\bi{britton12} Joseph W. Britton, Brian C. Sawyer, Adam C. Keith,
C.-C. Joseph Wang, James K. Freericks, Hermann Uys,
Michael J. Biercuk and John J. Bollinger, Nature {\bf 484}, 489 (2012).

\bi{islam13} R. Islam, C. Senkol, W. C. Campbell, S. Korenblit, J.
Smith, A. Lee, E. E. Edwards, C.-C. J. Wang, J. K. Freericks, and C. Monroe, Science {\bf 340}, 583 (2013).

\bi{richemere14}  P. Richerme, Z.-X. Gong, A. Lee, C. Senko, J. Smith, M.
Foss-Feig, S. Michalakis, A. V. Gorshkov, and C. Monroe,
Nature {\bf 511}, 198 (2014).


\bi{kitaev01} A. Kitaev, Physics-Uspekhi {\bf 44}, 131 (2001).
\bi{fulga11} I. C. Fulga, F. Hassler, A. R. Akhmerov, and C. W. J. Beenakker,
Phys. Rev. B {\bf 83}, 155429 (2011).
\bi{sau12} J. D. Sau and S. Das Sarma, Nature Comm. {\bf 3}, 964 (2012).
\bi{lutchyn11} R. M. Lutchyn and M. P. A. Fisher, Phys. Rev. B {\bf 84}, 214528
(2011).
\bi{degottardi11} W. DeGottardi, D. Sen, and S. Vishveshwara, New J. Phys. {\bf 13},
065028 (2011).
\bi{degottardi13} W. DeGottardi, D. Sen, and S. Vishveshwara, Phys. Rev. Lett.
{\bf 110}, 146404 (2013).
\bi{thakurathi13} M. Thakurathi, A. A. Patel, D. Sen, and A. Dutta, Phys. Rev. B
{\bf 88}, 155133 (2013).
\bi{degottardi13_1} W. DeGottardi, M. Thakurathi, S. Vishveshwara, and D. Sen,
Phys. Rev. B {\bf 88}, 165111 (2013).

\bi{alicea12} J. Alicea, Rep. Prog. Phys. {\bf 75}, 076501 (2012).

\bi{fu08}  L. Fu and C. L. Kane, Phys. Rev. Lett. {\bf 100}, 096407 (2008).


\bi{mourik12} V. Mourik et al., Science {\bf 336}, 1003 (2012).
\bi{deng12} M. T. Deng, C. L. Yu, G. Y. Huang, M. Larsson, P. Caroff, and
H. Q. Xu, Nano Lett. {\bf 12}, 6414 (2012).
\bi{das12} A. Das, Y. Ronen, Y. Most, Y. Oreg, M. Heiblum, and
H. Shtrikman, Nat. Phys.{\bf  8}, 887 (2012).
\bi{chang13}  W. Chang, V. Manucharyan, T. Jespersen, J. Nygard, and
C. Marcus, Phys. Rev. Lett. {\bf 110}, 217005 (2013).

\bi{flink13}  A. D. K. Finck, D. J. Van Harlingen, P. K. Mohseni, K. Jung,
and X. Li, Phys. Rev. Lett. {\bf 110}, 126406 (2013).

\bi{rainis13} Diego Rainis, Luka Trifunovic, Jelena Klinovaja, and Daniel Loss
Phys. Rev. B {\bf 87}, 024515 (2013).

\bibitem{lieb64} T. D. Schultz, D. C. Mattis, and E. H. Lieb, Rev. Mod. Phys. {\bf 36}, 856 (1964).

\bi{sachdev10} S. Sachdev, {\it Quantum Phase Transitions} (Cambridge University
Press, Cambridge, UK, 2011).



\bi{vodola14} D. Vodola, L. Lepori, E. Ercolessi, A.â€. V. Gorshkov, and G. Pupillo, Phys. Rev. Lett. {\bf 113}, 156402 (2014).

\bi{vodola_th} D. Vodola, \textit{Correlations and Quantum Dynamics of 1D Fermionic Models: New Results for the Kitaev Chain with Long-Range Pairing}, doctoral dissertation, University of Bologna, 2015, \href{url}{http://amsdottorato.unibo.it/id/eprint/6745}.

\bi{huang14} Z. Huang  and D. P. Arovas,  Phys. Rev. Lett. {\bf 113}, 076407 (2014).

\bi{vodola16} D. Vodola, L. Lepori, E. Ercolessi, and G. Pupillo, New J. Phys.
{\bf 18}, 015001 (2016).

\bi{viyuela16} O. Viyuela, D. Vodola, G. Pupillo, and M. A. Martin-Delgado,
Phys. Rev. B {\bf 94}, 125121 (2016).

\bi{lepori17} L. Lepori, A. Trombettoni, and D. Vodola, J. Stat. Mech. (2017) 033102.


\bi{viyuela15} O. Viyuela, A. Rivas and M. A.  Martin-Delgado,  2D Mater.
{\bf 2} 034006  (2015).

\bi{regemortel16}  M. Van Regemortel, D. Sels  and M. Wouters  Phys. Rev. A
{\bf 93} 032311 (2016).

\bi{lepori171} L. Lepori  and L. DellÕAnna   New J. Phys. {\bf 19},  103030 (2017).

\bi{giuliano18} D. Giuliano, S. Paganelli  and L. Lepori L Phys. Rev. B
{\bf 97},  155113 (2018).

\bi{cats18} P. Cats, A. Quelle, O. Viyuela, M. A. Martin-Delgado, and C. Morais Smith
Phys. Rev. B {\bf 97}, 121106 (R) (2018).


\bi{bhattacharya19} U. Bhattacharya, S. Maity, A. Dutta and D. Sen,  J. Phys. Condens. Matter {\bf 31}, 174003 (2019)
\bi{kosterlitz76} J. M. Kosterlitz, Phys. Rev. Lett. {\bf 37}, 1577 (1976).



\bi{bhattacharjee81} J. Bhattacharjee, S. Chakravarty, J. L. Richardson, and D. J.
Scalapino, Phys. Rev. B {\bf 24}, 3862 (1981).

\bi{luijten01} E. Luijten and H. Messingfeld,  Phys. Rev. Lett. {\bf 86} , 5305 (2001).

\bi{chaikin} P. M. Chaikin, and T. C. Lubensky, \textit{Principles of condensed matter physics}
(Cambridge University Press, Cambridge, England, 1995).

\bi{goldenfeld92}  N. Goldenfeld, \textit{ Lectures on phase transitions and the renormalization group} (Westview press, 1992)



\bi{anderson71} P. W. Anderson and G. Yuval, J. Phys. C 4, 607 (1971).

\bi{narayan99} O. Narayan, B. S. Shastry, J. Phys. A {\bf 32}, 1131 (1999); C. L. Kane and M. P. A. Fisher, Phys. Rev. B {\bf 46}, 15233 (1992); Y.-C. Tsai, J. Phys. A {\bf 31}, 2359 (1998).




\bi{dutta01}  A. Dutta and J.K. Bhattacharjee, Phys. Rev. B {\bf 64}, 184106 (2001).

\bi{sun17} G. Sun, Phys. Rev. A {\bf 96}, 043621 (2017).

\bi{fey16} S. Fey and K.P. Schmidt, Phys.  Rev.  B {\bf 94}, 075156 (2016).

\bi{kotliar83}  G. Kotliar, P.W. Anderson, and D.L. Stein, Phys. Rev. B {\bf 27}, 602 (1983).



\bi{dutta02} A Dutta,   Phys. Rev. B {\bf 65}, 224427 (2002).

\bi{aizenmann88} M. Aizenman, $et~al$,  J. Stat. Phys, {\bf 50}, 1 (1988).

\bi{dutta03} A. Dutta, Physica A {\bf 318}, 63 (2003).


\bi{defenu15} N. Defenu, A. Trombettoni, and A. Codello, Phys. Rev. E {\bf 92}, 052113 (2015).

\bi{defenu16} N. Defenu, A. Trombettoni, and S. Ruffo1, Phys. Rev. B {\bf 94}, 224411 (2016).

\bi{behan19} C. Behan, J. Phys. A {\bf 52}, 075401 (2019).

\bi{defenu17} N. Defenu, A. Trombettoni, and S. Ruffo1, Phys. Rev. B {\bf 96}, 104432 (2017).

\bibitem{alecce17} A. Alecce and L. Dell'Anna, Phys. Rev. B {\bf 95}, 195160 (2017).

\bi{altland97} A. Altland and M. R. Zirnbauer, Phys. Rev. B {\bf 55}, 1142 (1997).

\bi {berry84} M. V. Berry,   Proc. R. Soc. Lond. A {\bf 392}, 45  (1984).

\bi{zak89} J. Zak,   Phys. Rev. Lett. {\bf 62} 2747 (1989).


\bibitem{koffel12} T. Koffel, M. Lewenstien, and L. Tagliacozzo, Phys. Rev. Lett. {\bf 109}, 267203 (2012).

\bibitem{hauke10} P. Hauke, F. M. Cucchietti, A. M\"{u}ller-Hermes, M. C. Ba\~{n}uls, J. I. Cirac, and M. Lewenstien, New. J Phys. {\bf 12}, 113037 (2010).


\bibitem{muller12} D. Peter, S. M\"{u}ller, S. Wessel, and H. P. B\"{u}chler, Phys. Rev. Lett. {\bf 109}, 025303 (2012).

\bibitem{peschel03} I. Peschel, J. Phys. A: Math. Gen. {\bf 36}, L205-L208 (2003). 

\bibitem{eisert10} J. Eisert, M. Cramer and M.B. Plenio, Rev. Mod. Phys. {\bf 82}, 277 (2010).
\bibitem{holzhey94} C. Holzhey, F. Larsen and F. Wilczek, Nucl. Phys. B {\bf 424} 443-467 (1994).

\bibitem{vidal03} G. Vidal, J. I. Latorre, E. Rico, and A. Kitaev1, Phys. Rev. Lett. {\bf 90}, 22 (2003).

\bibitem{latorre04} J. I. Latorre, E. Rico, G. Vidal, Quant. Inf. Comput. {\bf 4}, 48-92 (2004).

\bibitem{calabrese04} P. Calabrese and J. Cardy, J. Stat. Mech. (2004) P06002.



\bi{lieb72} E. H. Lieb and D. W. Robinson, Communications in Mathematical Physics {\bf 28}, 251 (1972).



\bi{calabrese05} P. Calabrese and J. Cardy, J. Stat. Mech. (2005) {P04010}.

\bi{fagotti08} M. Fagotti and P. Calabrese, Phys. Rev. A {\bf 78}, 010306(R) (2008). 

\bi{rigol07} T. Kinoshita, T. Wenger, and D. S. Weiss, Nature {\bf 440},900 (2006); M. Rigol, V. Dunjko, V. Yurovsky, and M. Olshanii, Phys. Rev. Lett. {\bf 98}, 050405 (2007).

\bi{buyskikh16} A. S. Buyskikh, M. Fagotti, J. Schachenmayer, F. Essler, and
A. J. Daley, Phys. Rev. A {\bf 93}, 053620 (2016).

\bi{schachenmayer13} J. Schachenmayer, B. P. Lanyon, C. F. Roos, and A. J. Daley, Phys. Rev. X {\bf 3}, 031015 (2013).

\bi{hauke13} P. Hauke and L. Tagliacozzo, Phys. Rev. Lett. {\bf 111}, 207202 (2013).

\bi{worm13} M. van den Worm, B. C. Sawyer, J. J Bollinger and M. Kastner, New J. Phys. {\bf 15} (2013) 083007.

\bi{kastner15}  M. Kastner and M. van den Worm, Phys. Scr. {\bf T165} (2015) 014039.

\bi{mori19} T. Mori, J. Phys. A: Math. Theor. {\bf 52} (2019) 054001.

\bi{neyenhuis17}B. Neyenhuis, J. Zhang, P. W. Hess, J. Smith, A. C. Lee, P. Richerme, Z. Gong, A. V. Gorshkov and C. Monroe, Sci. Adv. 2017; {\bf 3}: e1700672.

\bi{ho18} W. W. Ho, I. Protopopov, and D. A. Abanin, Phys. Rev. Lett. {\bf 120}, 200601 (2018).

\bi{anirban17} A. Dutta and A. Dutta,  Phys. Rev. B {\bf 96}, 125113 (2017).

\bibitem{Kibble76} T. W. B. Kibble,
{J. Phys. A: Math. Gen.  {\bf 9}, 1387 (1976)};
{Phys. Rep. {\bf 67,} 183 (1980).}

\bibitem{Zurek96} W. H. Zurek,
{Nature (London) {\bf 317}, 505 (1985)};
{Acta Phys. Pol. B {\bf 24}, 1301 (1993)}; 
{Phys. Rep. {\bf 276}, 177 (1996).}

\bibitem{ZDZ05}\  W. H. Zurek, U. Dorner,  and P. Zoller, 
{Phys. Rev. Lett. {\bf 95}, 105701 (2005).}
%

\bibitem{Polkovnikov05}\  A. Polkovnikov, 
{Phys. Rev. B {\bf 72}, 161201(R) (2005).}

%
\bi{dziarmaga05} J. Dziarmaga, Phys. Rev. Lett. {\bf 95}, 245701 (2005).

\bibitem{Damski05}\  B. Damski,
{Phys. Rev. Lett. {\bf 95}, 035701 (2005).}
%
%

\bibitem{damski_zurek06}\  B. Damski and W. H. Zurek, 
{Phys. Rev. A {\bf 73}, 063405 (2005).}
\bibitem{mukherjee07} \ V. Mukherjee, U. Divakaran, A. Dutta, and D. Sen
{Phys. Rev. B {\bf 76}, 174303 (2007).}
\bibitem{divakaran08} \ U. Divakaran, A. Dutta, and D. Sen
{Phys. Rev. B {\bf 78}, 144301 (2008).}
\bibitem{sen08} \ D. Sen, K. Sengupta, and S. Mondal
{Phys. Rev. Lett. {\bf 101}, 016806 (2008).}

\bibitem{shreyoshi08} \ K. Sengupta, D. Sen, and S. Mondal
{Phys. Rev. Lett. {\bf 100}, 077204 (2008).}


\bibitem {dziarmaga10} J. Dziarmaga, Advances in Physics  {\bf 59}, 1063 (2010).
\bibitem{polkovnikov11} A. Polkovnikov, K. Sengupta, A. Silva, and M. Vengalattore, \textit{Colloquium: Nonequilibrium dynamics of closed interacting quantum systems}, Rev. Mod. Phys. {\bf 83}, 863 (2011).	
\bi{dutta15} A. Dutta, G. Aeppli, B. K. Chakrabarti, U. Divakaran, T. 
Rosenbaum, and D. Sen, \textit{Quantum Phase Transitions in Transverse Field 
	Spin Models: From Statistical Physics to Quantum Information} (Cambridge 
University Press, Cambridge, 2015).

\bi{landau} C. Zener, Proc. Roy. Soc. London Ser A {\bf 137}, 696 (1932); L. D.
Landau and E. M. Lifshitz, {\it Quantum Mechanics: Non-relativistic Theory}, 
2nd ed. (Pergamon Press, Oxford, 1965).

\bi{sei} S. Suzuki and M. Okada, in {\it Quantum Annealing and Related 
	Optimization Methods}, Ed. by A. Das and B. K. Chakrabarti (Springer-Verlag,
Berlin, 2005), p. 185.

\bi{vitanov} N. V. Vitanov and B. M. Garraway,  Phys. Rev. A {\bf 53}, 4288 (1996); N. V. Vitanov, {\it ibid.} {\bf 59}, 988 (1999).

\bi{functions} F. W. J. Olver, D. W. Lozier, R. F. Boisvert, and C. W. Clark, \textit{NIST Handbook of Mathematical Functions}
(Cambridge University Press, Cambridge, England, 2010); M. Abramowitz and I. A. Stegun, \textit{Handbook of Mathematical Functions}(Dover, 1964).

\bi{jaschke17} D. Jaschke, K.  Maeda, J. DWhalen and M. L. Wall and L. D Carr, New J. Phys. {\bf 19}  033032 (2017).



\bibitem{nandy18} S. Nandy, K. Sengupta, and A. Sen, J. Phys. A: Math. Theor. {\bf 51} 334002 (2018).

\bi{sen16} A. Sen, S. Nandy S and K. Sengupta, Phys. Rev. B {\bf 94} 214301 (2016).


\bi{heyl13} M. Heyl, A. Polkovnikov, and S. Kehrein, Phys. Rev. Lett., {\bf 110}, 135704 (2013).

\bi{pollmann10} F. Pollmann, S. Mukerjee, A. G. Green, and J. E. Moore, Phys. Rev. E {\bf 81}, 020101(R) (2010).

\bi{heyl14} M. Heyl, Phys. Rev. Lett., {\bf 113}, 205701 (2014).

\bi{vajna14} S. Vajna, and B. Dora,  Phys. Rev. B {\bf 89}, 161105(R) (2014).

\bi{heyl15} M. Heyl, Phys. Rev. Lett., {\bf 115}, 140602 (2015) .

\bi{sharma16} S. Sharma, U. Divakaran, A. Polkovnikov, and A. Dutta, Phys. Rev. B {\bf 93}, 144306 (2016).

\bi{divakaran16} U. Divakaran, S. Sharma, and A. Dutta, Phys. Rev. E {\bf 93}, 052133 (2016).

\bi{karrasch13} C. Karrasch and D. Schuricht,  Phys. Rev. B, {\bf 87}, 195104 (2013).

\bi{kriel14} N. Kriel, C. Karrasch, and S. Kehrein, Phys. Rev. B {\bf 90}, 125106 (2014).

\bi{sharma15} S. Sharma, S. Suzuki, and A. Dutta,  Phys. Rev. B {\bf 92}, 104306 (2015).


 \bibitem{jurcevic16}  P. Jurcevic, H. Shen, P. Hauke, C. Maier, T. Brydges, C. Hempel, B. P. Lanyon, M. Heyl, R. Blatt, C. F. Roos,  arXiv:1612.06902 (2016).
%
\bi{flaschner16} N. Fl\"aschner, D. Vogel, M. Tarnowski, B, S. Rem, D.-S. L\"uhmann, M. Heyl, J.  Budich, L. Mathey, K. Sengstock, C. Weitenberg,
arXiv:1608.05616 (2016).

\bi{budich15} J. C. Budich and  M. Heyl,  Phys. Rev. B {\bf 93}, 085416 (2016). 

\bi{schmitt15} M. Schmitt and S. Kehrein, Phys. Rev. B {\bf 92}, 075114 (2015).

\bi{vajna15} S. Vajna and B. Dora, Phys. Rev. B {\bf 91}, 155127 (2015).

\bi{utso16} U. Bhattacharya and A. Dutta, Phys. Rev. B {\bf 95}, 184307 (2017).

\bi{utso17} U. Bhattacharya and A. Dutta, Phys. Rev. B {\bf 96}, 014302  (2017). 

\bi{bhattacharya17} U. Bhattacharya, S. Bandyopadhyay and A. Dutta, Phys. Rev. B {\bf 96}, 180303 (R) (2017). 

\bi{heyl17} M. Heyl and J.  C. Budich, Phys. Rev. B {\bf 96}, 180304 (R) (2017). 

\bi{bandyopadhyay17} S. Bandyopadhyay, S. Laha, U. Bhattacharya and A. Dutta, Scientific Reports {\bf 8}, 11921 (2018).

\bi{spin_model_dpt}B \v Zunkovi\v c, M. Heyl, M. Knap, A. Silva, Phys. Rev. B {\bf 96}, 134313 (2017).

\bi{zuncovic16} B. Zunkovic,  A/ Silva A, M. Fabrizio, Philos Trans A Math Phys Eng Sci. {\bf 13}, 374 (2016).


\bi{halimeh17} J. C. Halimeh, V. Zauner-Stauber, I. P. McCulloch, I. de Vega, U. Schollwšck, and M. Kastner, Phys. Rev. B {\bf 95}, 024302 (2017).

\bi{halimeh171} J. C. Halimeh and V. Zauner-Stauber, Phys. Rev. B {\bf 96}, 134427 (2017)

\bi{homrighausen17} I. Homrighausen, N. O. Abeling, V. Zauner-Stauber, and J. C. Halimeh, Phys. Rev. B {\bf 96}, 104436 (2017).



\bi{piccitto19} G. Piccitto, B.  Zunkovic, and A. Silva, arXiv:1906.00691 (2019).

\bi{rajak14} A. Rajak and A. Dutta, Phys. Rev. E {\bf 89}, 042125 (2014).

\bi{patel13} A. A. Patel, S. Sharma, and A. Dutta, Eur. Phys. J. B {\bf 86}, 367 (2013).

\bi{sacramento14} P. D. Sacramento, Phys. Rev. E {\bf 90} 032138 (2014).

\bi{mishra18} U. Mishra, R. Jafari, A.  Akbari, arXiv:1810.06236 (2018).

\bi{kastner10} M.  Kastner,
Phys. Rev. Lett. {\bf 104}, 240403  (2010).

\bi{kastner11} M. Kastner, Phys. Rev. Lett. {\bf 106}, 130601 (2011).

\bi{sweke19} R. Sweke, J. Eisert,  and M.  Kastner, arXiv: 1906.00791 (2019).


\bi{saadatmand18} S.N. Saadatmand, S.D. Bartlett, and I.P. McCulloch, Phys. Rev. B {\bf 97}, 155116 (2018).

\bi{fey19} S. Fey, S.C. Kapfer, K.P. Schmidt, Phys.
Rev. Lett.  {\bf 122}, 017203 (2019).

\bi{pientka17} F. Pientka, A. Keselman, E. Berg, A. Yacoby, A. Stern,
B.I. Halperin, Phys. Rev. X {\bf 7}, 021032 (2017).

\bi{liu18} D. T. Liu, J. Shabani, and A. Mitra, Phys. Rev. B {\bf 97}, 235114 (2018).

\bi{shiba68}  H. Shiba, Prog. Theor. Phys. {\bf 40}, 435 (1968).

\bi{hoffman16} S. Hoffman, J. Klinovaja, and D. Loss, Phys. Rev. B {\bf 93}, 165418 (2016).

\bi{menard17} G. C. Menard, Nat. Comm.  {\bf 8},  2040 (2017).

\bi{viyuela18} O. Viyuela, L. Fu, M. A. Martin-Delgado,  Phys. Rev. Lett., {\bf 120}, 017001 (2018).

\end{thebibliography}
\end{document}